\documentclass[preprint,12pt]{elsarticle}

\makeatletter
\newif\iffinalversion\finalversionfalse
\newif\ifpreprintversion\preprintversionfalse
\newif\ifreviewversion\reviewversionfalse
\newif\ifwithtoc\withtocfalse
\newif\ifwithlinenumbers\withlinenumbersfalse
\newif\ifsectioninput\sectioninputfalse
\makeatother

\preprintversiontrue\finalversionfalse 

\iffinalversion

  \withtocfalse          
  \withlinenumbersfalse  
  \sectioninputtrue
  \long\def\xblue#1{}   

  \def\sout#1{}  
\else

  \ifreviewversion
     
     \long\def\xblue#1{}
     
     \withtocfalse
     \withlinenumbersfalse  
     \sectioninputtrue
  \else
     \withtoctrue
     \ifpreprintversion
       \withlinenumbersfalse
       
       \long\def\xblue#1{}
       
     \else 
       \withlinenumberstrue
       
       \long\def\xblue#1{{\color{blue}{#1}}}
       
     \fi
  \fi

  \usepackage[normalem]{ulem}
\fi
\def\prgname#1{#1} 

\usepackage{graphicx}
\usepackage{amssymb}
\usepackage{amsmath}
\usepackage{amsthm}
\usepackage{eurosym}
\usepackage{hyperref}
\usepackage{color}
\usepackage{url}

\ifwithlinenumbers
\usepackage{lineno}
\fi

\makeatletter
\newif\ifwithtwocolumns\withtwocolumnsfalse
\if@twocolumn\withtwocolumnstrue\fi 
\makeatother
\iffinalversion\withtwocolumnstrue\fi

\def\hc#1{\leavevmode\hbox to \hsize{\hss #1\hss}\leavevmode}

\ifwithtwocolumns
  \def\includefigure#1{\hc{\resizebox{\columnwidth}{!}{\includegraphics{#1}}}}
\else
  \def\includefigure#1{\hc{\resizebox{11.6cm}{!}{\includegraphics{#1}}}}
\fi

\journal{Astroparticle Physics}

\begin{document}

\begin{frontmatter}

\title{Monte Carlo design studies for the Cherenkov Telescope Array}

\author[inst-mpik,inst-hu]{K.~Bernl{\"o}hr\corref{cor1}}
       \ead{Konrad.Bernloehr@mpi-hd.mpg.de}
\author[inst-copernicus]{A.~Barnacka}
\author[inst-llr,inst-apc]{Y.~Becherini}
\author[inst-ifae]{O.~Blanch Bigas}
\author[inst-mppmu,inst-ciemat]{E.~Carmona}
\author[inst-mppmu]{P.~Colin}
\author[inst-desyz,inst-argonne]{G.~Decerprit}
\author[inst-torino]{F.~Di~Pierro}
\author[inst-madrid-comp]{F.~Dubois}
\author[inst-stockholm-phys,inst-stockholm-klein]{C.~Farnier}
\author[inst-slac-kavli]{S.~Funk}
\author[inst-mpik]{G.~Hermann}
\author[inst-leicester]{J.A.~Hinton}
\author[inst-columbia]{T.B.~Humensky}
\author[inst-llr]{B.~Kh\'elifi}
\author[inst-mpik]{T.~Kihm}
\author[inst-annecy]{N.~Komin}
\author[inst-lsw,inst-obspm]{J.-P.~Lenain}
\author[inst-desyz]{G.~Maier}
\author[inst-ifae]{D.~Mazin}
\author[inst-saclay]{M.C.~Medina}
\author[inst-ifae]{A.~Moralejo}
\author[inst-durham]{S.J.~Nolan}
\author[inst-leicester,inst-leeds]{S.~Ohm}
\author[inst-mpik]{E.~de~O{\~n}a Wilhelmi}
\author[inst-leeds,inst-mpik]{R.D.~Parsons}
\author[inst-desyz,inst-hu]{M.~Paz Arribas}
\author[inst-ieec]{G.~Pedaletti}
\author[inst-apc]{S.~Pita}
\author[inst-desyz]{H.~Prokoph}
\author[inst-durham]{C.B.~Rulten}
\author[inst-hu]{U.~Schwanke}
\author[inst-desyz]{M.~Shayduk}
\author[inst-ifae]{V.~Stamatescu}
\author[inst-torino]{P.~Vallania}
\author[inst-lupm,inst-desyz]{S.~Vorobiov}
\author[inst-desyz]{R.~Wischnewski}
\author[inst-icrr]{T.~Yoshikoshi}
\author[inst-obspm]{A.~Zech}
\author[]{for the CTA Consortium}

\cortext[cor1]{Corresponding author}

\address[inst-mpik]{Max-Planck-Institut f{\"u}r Kernphysik, P.O. Box 103980, 
        D-69029 Heidelberg, Germany}
\address[inst-hu]{Institut f{\"u}r Physik, Humboldt-Universit{\"a}t zu Berlin, 
        Newtonstr. 15, D-12489 Berlin, Germany}
\address[inst-copernicus]{Nicolaus Copernicus Astronomical Center, Polish Academy
        of Sciences, ul. Bartycka 18, 00-716 Warsaw, Poland}
\address[inst-llr]{Laboratoire Leprince-Ringuet, Ecole Polytechnique, CNRS/IN2P3,
        F-91128 Palaiseau, France}
\address[inst-apc]{Astroparticule et Cosmologie (APC), CNRS, Universit{\'e} Paris 7
        Denis Diderot, Paris Cedex 13, France}
\address[inst-ifae]{IFAE, Edifici Cn., Campus UAB, E-08193 Bellaterra, Spain}
\address[inst-mppmu]{Max-Planck-Institut f{\"u}r Physik, F{\"o}hringer Ring 6, 
        D-80805 M{\"u}nchen, Germany}
\address[inst-ciemat]{Centro de Investigaciones Energ{\'e}ticas, Medioambientales y
        Tecnol{\'o}gicas (CIEMAT), Madrid, Spain}
\address[inst-desyz]{DESY, Platanenallee 6, D-15738 Zeuthen, Germany}
\address[inst-argonne]{Argonne National Laboratory, 9700 S.~Cass Avenue, Argonne, 
	    IL~60439, USA}
\address[inst-torino]{Osservatorio Astrofisico di Torino dell'Istituto Nazionale
        di Astrofisica, Corso Fiume 4, I-10133 Torino, Italy}
\address[inst-madrid-comp]{Universidad Complutense, E-28040 Madrid, Spain}
\address[inst-stockholm-phys]{Department of Physics, Stockholm University, 
        AlbaNova, SE-106~91 Stockholm, Sweden}
\address[inst-stockholm-klein]{The Oskar Klein Centre for Cosmoparticle Physics, 
        AlbaNova, SE-106~91 Stockholm, Sweden}
\address[inst-slac-kavli]{Kavli Institute for Particle Astrophysics and Cosmology, 
        SLAC, Stanford, CA~94025, USA}
\address[inst-leicester]{Department of Physics and Astronomy, The University of Leicester,
        Leicester, LE1~7RH, United Kingdom}
\address[inst-columbia]{Physics Department, Columbia University, New York, NY 10027, USA}
\address[inst-annecy]{LAPP, Universit{\'e} de Savoie, CNRS/IN2P3, 
        F-74941 Annecy-le-Vieux, France}
\address[inst-lsw]{Landessternwarte, Universit\"at Heidelberg, K\"onigstuhl, 
        D-69117 Heidelberg, Germany}
\address[inst-obspm]{LUTH, Observatoire de Paris, CNRS, Universit{\'e} Paris Diderot, 
        5~place Jules Janssen, F-92190 Meudon, France}
\address[inst-saclay]{CEA Saclay, DSM/IRFU, F-91191 Gif-Sur-Yvette Cedex, France}
\address[inst-durham]{University of Durham, Department of Physics, 
        South Road, Durham DH1~3LE, United Kingdom}
\address[inst-leeds]{University of Leeds, School of Physics and Astronomy, 
        Leeds LS2~9JT, United Kingdom }
\address[inst-ieec]{Institut de Ci{\`e}ncies de l'Espai (IEEC-CSIC), Campus UAB, Torre C5,
        E-08193 Barcelona, Spain}
\address[inst-lupm]{LUPM, UMR~5299 Universit{\'e} Montpellier II \&
        IN2P3/CNRS, F-34095 Montpellier, France}
\address[inst-icrr]{Institute for Cosmic Ray Research, The University of Tokyo, Kashiwa,
        Chiba 277-8582, Japan}

\begin{abstract}
The Cherenkov Telescopes Array (CTA) is planned as the
future instrument for very-high-energy (VHE) $\gamma$-ray
astronomy with a wide energy range
of four orders of magnitude and an 
improvement in sensitivity compared to current instruments 
of about an order of magnitude.
Monte Carlo simulations are a crucial tool in the design of CTA. 
The ultimate goal of these
simulations is to find the most cost-effective solution for 
given physics goals and thus sensitivity goals or to find,
for a given cost, the solution best suited for different types of targets
with CTA. Apart from uncertain component cost estimates, the main
problem in this procedure is the dependence on a huge number
of configuration parameters, both in specifications of individual telescope
types and in the array layout. This is addressed by simulation
of a huge array intended as a superset of many different 
realistic array layouts, and also by simulation of array subsets
for different telescope parameters. Different analysis methods
-- in use with current installations and extended (or developed specifically)
for CTA -- are applied to the simulated data sets for deriving
the expected sensitivity of CTA. In this paper we describe
the current status of this iterative approach to optimize the
CTA design and layout.

\end{abstract}


\end{frontmatter}

\ifwithtoc
\clearpage
\tableofcontents
\listoffigures 
\fi

\ifwithlinenumbers
\linenumbers
\fi


\ifsectioninput

\section{Introduction}
\label{sec:introduction}

The concept of the Cherenkov Telescope Array (CTA) \cite{cta-www,cta-intro} is based on a
straightforward expansion of current imaging atmospheric Cherenkov
telescope (IACT) arrays for very high energy (VHE) gamma-ray
astrophysics such as H.E.S.S., MAGIC, and VERITAS 
\cite{hess-www, magic-www, veritas-www}. It aims to improve
the current sensitivity by an order of magnitude and extend the
current sensitive energy region to lower and higher energies covering
a range of about four orders of magnitude. In order to study
astrophysical objects in such a wide energy range, CTA will consist of
at least three different sized telescopes: the Large Size Telescopes (LSTs,
$\sim$24~m aperture) will record showers with energies as low as
$\sim$20~GeV, the Medium Size Telescopes (MSTs, $\sim$12~m
aperture, later to be supplemented by SC-MSTs of 9~m aperture
with Schwarzschild-Couder (SC) optics \cite{SCoptics}) will operate in the $\sim$1~TeV range, and the Small Size
Telescopes (SSTs, $\sim$4--7~m aperture) are optimised for high
energies up to more than $100$~TeV. 
This design scenario is schematically
shown in Figure~1 of the CTA Design Concepts \cite{cta2010} and 
possible implementations
are illustrated in Section~\ref{sec:prod1-config} of the current paper. 
The performance of an IACT array is also determined by quantities like the
angular resolution, energy resolution or sensitive field of view
(f.o.v.). These performance estimators depend on a large number of
technical and design parameters within a cost envelope. The role of
the Monte Carlo (MC) Work Package (WP) of CTA is to optimize the array
configuration in this parameter space using MC simulations, given
scientific requirements as suggested by the Physics (PHYS) WP.

Our extensive simulation studies to optimize the CTA design are also
motivated by several earlier simulation studies, in the following separated
into three energy regimes. In the {\bf low energy regime}, from 100~GeV
down to a few GeV in the most extreme proposals, 
it is essential to collect more Cherenkov
photons from a gamma-ray shower with a very large aperture telescope of the
20--30~m scale and/or at a very high altitude of $\gtrsim$4000~m. 
ECO-1000 (European Cherenkov Observatory, with a mirror
surface of 1000~m$^2$) has been the most extreme proposal
in terms of mirror area of a single telescope \cite{bai2004}. 
On the basis of the former
ideas, a large single telescope of 34~m aperture with high
quantum efficiency photon detectors in the focal plane for a further
light gain was proposed. 
The large proposed aperture and efficiency raises
major noise problems. Simulation studies for such a huge telescope show that
night sky background (NSB) light collected by the large reflector
causes high accidental trigger rates and a smaller pixel size is
preferable to reduce this effect. In addition, Cherenkov light from single
secondary muons arriving at large distances of $\gtrsim$40~m results in
gamma-ray-like images which cannot be rejected completely by a single
telescope. The latter noise due to single muons can be eliminated by
stereoscopic IACT arrays.
An array of 30~m telescopes was proposed \cite{kon2005} for a 10~GeV threshold
at 1800~m altitude.
Another proposed project, 5@5, was intended as an array of
4--5 IACTs of $\sim$20~m aperture, aiming at a threshold as low as 
$\sim$5~GeV gamma rays at a very high altitude site of about 5~km a.s.l.\
\cite{aha2001}. This low threshold at high altitude can be achieved
because, being closer to shower maximum, the Cherenkov light is
less diluted -- but at the cost of smaller effective areas and
more difficult gamma-hadron separation when electrons in most gamma-ray showers
reach the telescopes.

Cherenkov telescopes should focus light coming from a distance $d$
(typical distance of the average shower maximum) onto the pixels.
If defocusing by half of a pixel diameter $p$ is acceptable,
the depth of field is then from about $d/(1+pd/(2fD))$ to $d/(1-pd/(2fD))$
(or infinity if $pd\ge 2fD$), with focal length $f$ and telescope diameter $D$.
While the depth of field of small Cherenkov telescopes encompasses
most showers entirely, a problem with large telescopes -- particularly severe
at high altitude -- is the very limited depth of field, unable to
focus all parts of shower images at the same time \cite{hof2001}.
The useful size of large Cherenkov telescopes is thus not 
just limited by their cost and CTA does not plan to build
extremely large telescopes.

In the {\bf medium energy regime} from about 100 GeV to some 10~TeV, 
CTA aims to improve the
sensitivity by stereoscopic observations using several IACTs
simultaneously. A previous simulation study of a dense IACT array
(33.3~m telescope spacing) shows that the angular resolution $\sigma$
depends on the number of telescopes $N$ used in the reconstruction and
is improved as $\propto${}$N^{-1/2}$ up to $N \sim 50$ \cite{hof1999}. The
point-source sensitivity is limited by background fluctuations in this
energy regime and approximately proportional to $\sigma A^{-1/2}
Q^{-1}$, where $A$ is the effective area and 
$Q=\epsilon_\gamma/\sqrt{\epsilon_{\rm bg}}$ is the quality
factor usually defined in terms of the gamma-ray and
background cut efficiencies $\epsilon_\gamma$ and
$\epsilon_{\rm bg}$, respectively. 
There is a trade-off between $\sigma$ and $A$ since the number
of available telescopes is limited by the cost. However, high
telescope-multiplicity observations also give better background
rejection and thus a larger $Q$.

In the {\bf high energy regime} beyond some 10~TeV, 
an extensive IACT array called TenTen has
been proposed aiming to achieve a 10~km$^2$ effective area at
energies greater than 10~TeV \cite{row2008}. It consists of
relatively small IACTs of 3--5~m aperture located with an
inter-telescope spacing of $\gtrsim$300~m. This sparse array design,
which was first suggested by Plyasheshnikov et al.\ (2000)
\cite{ply2000}, enables less IACTs to expand the effective area
cost-effectively. The IACTs are also required to have a larger f.o.v.\ diameter of
$\gtrsim${}$8^\circ$ as they detect Cherenkov photons at large core
distances beyond the plateau area of the Cherenkov light pool (radius
$\sim$120~m). Also there, the Cherenkov photon density sharply
drops with increasing the core distance, with consequences on the
energy threshold and also on
the accuracies of determining the arrival direction and the core
position (see \cite{col2009} on the optimization of an IACT array).
The stereoscopic reconstruction can also be complemented using the time
gradient analysis \cite{sta2011}. Also, the simulation study by de la
Calle P\'{e}rez \& Biller (2006) indicates that the sensitivity can be
improved by more than three times above 300~GeV and significantly more
than this above 10~TeV, only by using a 
wide f.o.v.\ camera of
$10^\circ$ diameter with a conventional telescope spacing\footnote{The
authors also utilized a time gradient cut and cuts with shower
reconstruction parameters (energy and shower maximum height) to
improve the sensitivity.} \cite{del2006}.

The above mentioned ideas and many technical aspects should be 
considered in the Monte Carlo
simulation study for CTA. The optimization of the array
configuration within such a large parameter space is thus a challenging
task. Before the detailed CTA design study, preliminary simulations
have been carried out with some homogeneous and graded array
configurations, focusing mainly on the low to medium energy range
\cite{ber2008a}, which demonstrated that the CTA goal sensitivity can
be achieved within an anticipated budget.

So far, only single mirror optics such as the parabolic or
Davies-Cotton (DC) designs have been used for IACTs, although
secondary mirrors are commonly used in optical telescopes in order to
get much better spot sizes correcting the spherical and coma
aberrations. The AGIS (Advanced Gamma-ray Imaging System) group
\cite{agis}, now part of CTA, proposed an advanced idea utilizing a two
mirror optics called the Schwarzschild-Couder optics for IACTs
\cite{SCoptics}. With these optics, IACTs can have a very large f.o.v.\ up
to $\sim${}$15^\circ$ without significant degradation of the spot size
and therefore increase survey capabilities with a good angular
resolution. These optics possibly give some other advantages: the
physical pixel size is more compact than that of the single mirror
optics and cost-effective photon sensors such as multi-anode
photomultiplier tubes (MAPMTs) or silicon photomultipliers (SiPMs) can
be used for the camera. 
Moreover, the physical telescope
length along the optical axis gets shorter and the cost of the
telescope structure can be reduced significantly. 
On the other hand, requirements on the optical quality get
more ambitious. A portion of CTA (mainly US groups, following up
on the AGIS project) aims to complement the baseline CTA design with a
sub-array of 9.5~m diameter SC-MSTs.
The SC design is also attractive for the SST array because of its 
wide f.o.v.\ capability, which is needed for the sparse telescope array concept.
Two different optical designs for 4~m SC-SSTs and two different
camera concepts for them are under development.

In this paper, we describe our detailed simulation study for CTA
carried out after \cite{ber2008a}. In Section~\ref{sec:tools}, details of the
software tools used for our simulation study are summarized. Our
simulation work is an integration of several activities in this field
and various data analysis methods are simultaneously
considered and developed. We describe details of the analysis methods
in Section~\ref{sec:baseline-analysis} and \ref{sec:alt_methods}. 
The interface to and tools for the physics program of CTA are
summarized in Section~\ref{sec:mc2phys}.
The common array configurations used in the mass
production of simulation data are presented in 
Section~\ref{sec:prod1-config}. Our current
results of performance estimations for candidate arrays with the
baseline analysis are described in 
Section~\ref{sec:baseline-performance} and compared with those
of the other analyses in Section~\ref{sec:alt_ana_comparison}. 
We describe future directions of
our study in Section~\ref{sec:FutureHardware} and conclude in 
Section~\ref{sec:conclusion}.

\section{Monte Carlo simulation tools}
\label{sec:tools}


\subsection{Air shower simulation}
\label{subsec:airshower}

The first step in the generation of very high energy (VHE) $\gamma$-ray events or
background (cosmic ray) events for CTA is the simulation of the extensive
air shower, i.e.\ the cascade of secondary particles developing in the
atmosphere. Several codes exist for the detailed, three-dimensional MC
simulation of air showers for different primary particles. A MC
simulation of the shower development, rather than an analytical solution
of the cascade equations, is necessary to correctly account for
statistical fluctuations between showers. The main challenge of these
simulations is the correct treatment of hadronic interactions, which
play a central role in the development of air showers triggered by
cosmic rays. Phenomenological models are used to extrapolate
cross-sections beyond the energy regime and scattering angles
accessible to accelerator experiments. 

The air shower generator \prgname{CORSIKA} \cite{corsika} has been chosen as a
standard tool for CTA simulations. This publicly available, open-source
code is used by all the current  IACT arrays and represents a standard
tool in the wider astroparticle physics community. While electromagnetic
interactions in \prgname{CORSIKA} are treated by an adapted version of the 
\prgname{EGS4} code~\cite{egs4}, a choice is given between several hadronic 
interaction models. The ``IACT/ATMO'' package \cite{ber2008} facilitates
the simulation of the Cherenkov light flux for a chosen configuration of
telescope positions and dish sizes. The flux of Cherenkov photons from a
simulated air shower is collected for each pre-defined telescope
position and dish size and serves as input for further processing with a
telescope simulation package.

The generation of proton-induced showers, needed for background
estimations, largely dominates the CPU time and requires substantial
amounts of memory and disk space. The large volume of simulations needed
to investigate all the different configurations for the array has
motivated the use of the EGEE/EGI (Enabling Grids for 
E-sciencE/European Grid Initiative) Computing Grid for the massive
production of shower and detector simulations, in addition to
the use of local computing resources.  Within the Grid virtual
organisation for CTA, 14 computing centres located at collaborating
institutes provide computing power and storage. To save disk space and
CPU time, the output of \prgname{CORSIKA} is usually piped directly into several instances
of the telescope simulation software to generate the response of
different array configurations in parallel, without writing the large
\prgname{CORSIKA} output files to disk. At peak times of MC production, up to 2000
simulation jobs can run in parallel. The large distributed data storage
space (several 100 TB on disks and tapes) makes it also possible to
temporarily store \prgname{CORSIKA} files for later reprocessing, 
to compare for example the same showers as seen with
different nightsky background or for different telescope implementation details.
It is foreseen
to equally perform the further processing and analysis of the simulated
data on the Computing Grid in the near future.  In addition, massive
simulations have been carried out on local CPU clusters at several CTA
member institutes and the data are provided for download and were
used for the baseline analysis as well as for some of the alternative analyses.

The Cherenkov light production from $\gamma$-ray events simulated with
\prgname{CORSIKA} has been cross-checked against another air shower generator
(\prgname{KASCADE-C$++$}) currently in use within the H.E.S.S. collaboration. This
code had been developed by the ARTEMIS-Whipple, CAT, and H.E.S.S.
collaborations, based on the original \prgname{KASCADE} code~\cite{kascade}. The
Cherenkov light profile generated by the two air shower codes agrees to
within $\sim$5\% (cf. Fig.~\ref{fig:light_cors_kas}), resulting in
consistent telescope trigger rates and photo-electron (p.e.)
distributions in the camera.  

\begin{figure}[htbp]
\begin{minipage}[t]{0.49\columnwidth}
\mbox{}\\
\centerline{\includegraphics[width=\columnwidth]{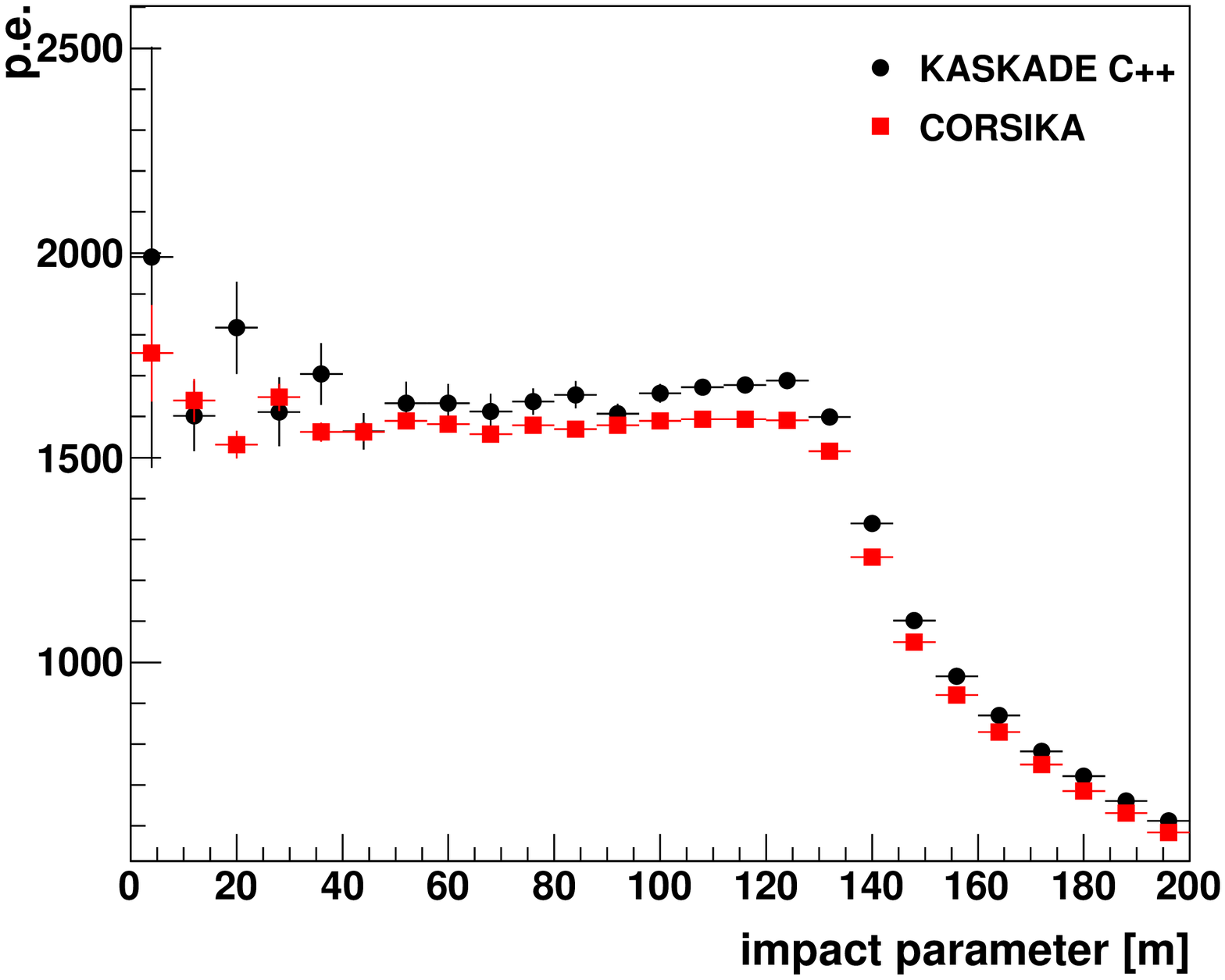}}
\end{minipage}
\hfill
\begin{minipage}[t]{0.49\columnwidth}
\mbox{}\\
\centerline{\includegraphics[width=\columnwidth]{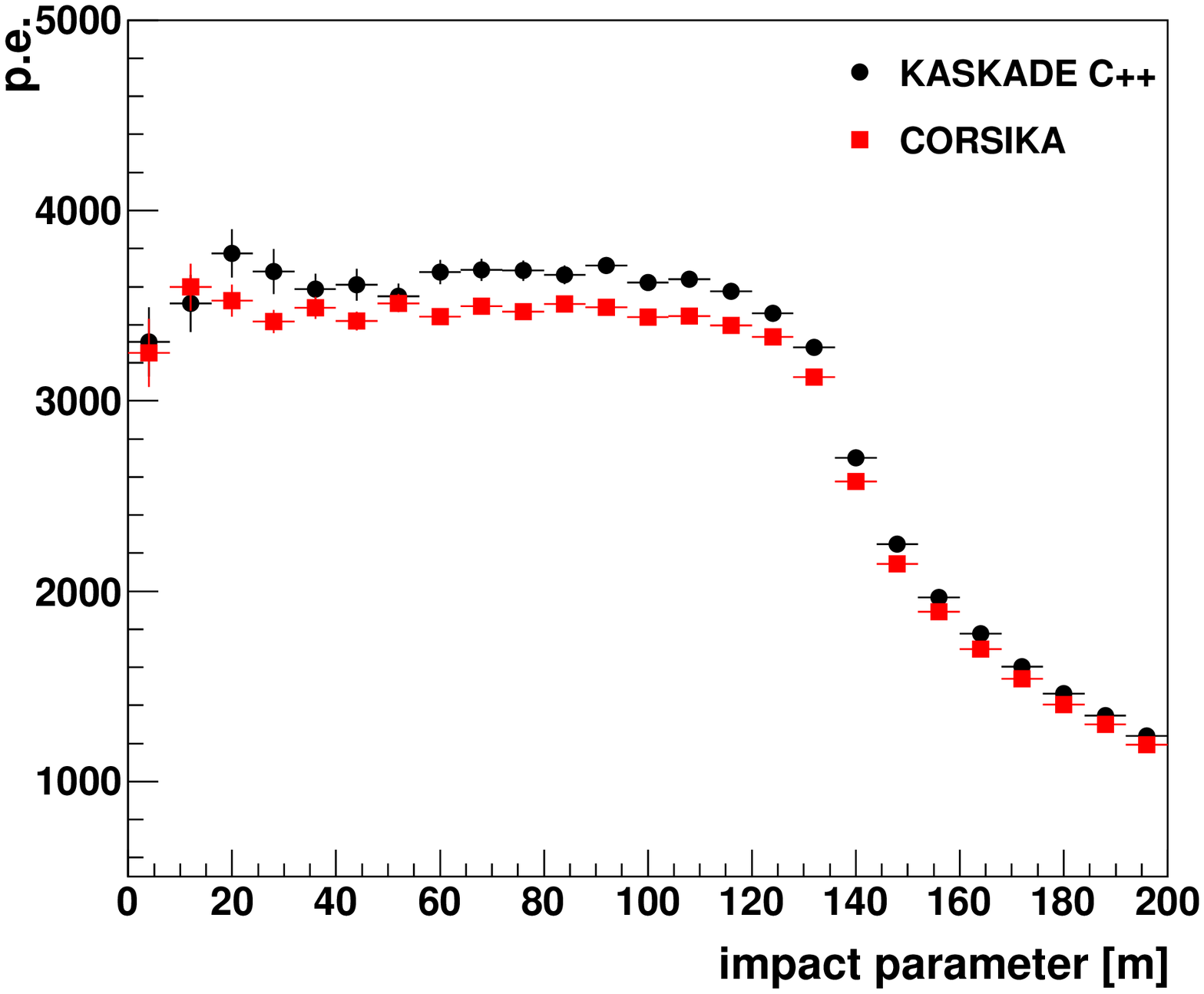}}
\end{minipage}
 \caption[Cherenkov profiles CORSIKA / KASKADE-C$++$]
 {Simulated Cherenkov light profiles (in p.e.\ collected by the camera on a 106~m$^2$ H.E.S.S.
telescope \cite{hess-cam,hess-tel}) for vertical $\gamma$-ray showers at 500 GeV (left) and 1 TeV
(right) for \prgname{CORSIKA} and \prgname{KASKADE-C$++$}, as a function of the impact
parameter. No atmospheric extinction of the Cherenkov light has been applied.}
\label{fig:light_cors_kas}
\end{figure}


\begin{figure}[htbp]
\includefigure{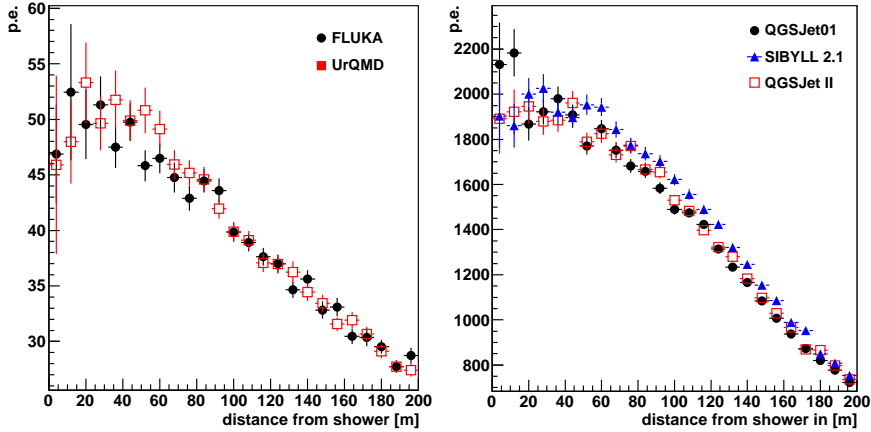}
\caption[Cherenkov light profiles]{Comparison of the simulated
Cherenkov light profiles for vertical proton-induced showers generated
by \prgname{CORSIKA} with different hadronic interaction models. The profiles of
p.e.\ collected by a H.E.S.S. camera \cite{hess-cam,hess-tel} 
for \prgname{FLUKA} and \prgname{UrQMD} at 100 GeV are shown in the left panel. 
In the right panel, the high-energy interaction models
QGSJet-01c, SIBYLL, and QGSJet-II are compared
for showers induced by 1.0~TeV protons,
all using URQMD for low-energy interactions.}
\label{fig_had_models}
\end{figure}

The simulation of the cosmic ray background is subject to our still
limited knowledge of hadronic interaction processes at very high
energies.  Detailed comparisons of the different interaction models
available in \prgname{CORSIKA} can be found in~\cite{corsika} and an evaluation of
the systematic uncertainties is given by~\cite{par2011}. The impact on
the Cherenkov light profile has been studied for some of the most
commonly used interaction models for low and high proton energies. 
There are no significant differences
between the low energy ($<$80 GeV) models \prgname{FLUKA}~\cite{fluka} and
\prgname{UrQMD}~\cite{urqmd}. The known discrepancy between the high-energy models
\prgname{QGSJet-01}~\cite{qgsjet01}, \prgname{QGSJet-II}~\cite{qgsjet2a, qgsjet2b} and
\prgname{SIBYLL~2.1}~\cite{sibyll}, which apply different extrapolations of
interaction parameters to high energies, leads only to a small
uncertainty of about 5\% in the Cherenkov light profile at 1 TeV. 
Examples of light profiles are shown for comparison in Fig.~\ref{fig_had_models}.
\prgname{UrQMD} and \prgname{QGSJet-II} had been chosen for the massive 
background simulations for CTA.


\subsection{Cherenkov telescope simulation}
\label{subsec:telsim}
 
The simulation of the detector response includes the optical ray-tracing
of the photons from the mirror to the photomultiplier tubes in the camera, 
the electronics and the digitization of the signals, as well as the trigger system.
Noise from
the night-sky background and from the electronics need to be added to the
signal as well. The telescope simulation package
\prgname{sim\_telarray}~\cite{ber2008}, based on software developed for HEGRA and
now in use by the H.E.S.S. collaboration, is employed for the massive
simulations for CTA. In general, \prgname{sim\_telarray} requires only a small
fraction of the CPU time needed for shower simulations. Once an air
shower has been simulated by \prgname{CORSIKA} and processed through 
sim\_telarray, only the final output is written to disk.

The simulation of the CTA instrumental response has been cross-checked
with the \prgname{SMASH} software \cite{smash}, 
which is also used within the H.E.S.S.
collaboration, and with the code used by the MAGIC collaboration
\cite{magicMC}. The agreement with sim\_telarray is generally very good.
As an example, $\gamma$-ray induced air showers at 1~TeV and with a
zenith angle of 30$^{\circ}$ have been generated with \prgname{CORSIKA} at a
distance of about 150 m from  the center of a H.E.S.S. type
four-telescope array. 
The distribution of the average
signals for 100 simulated showers processed by the two programs is
shown in Fig.~\ref{fig:adc_distribution}.  All of the showers used to
generate Fig.~\ref{fig:adc_distribution} have an impact parameter to the
simulated telescope of 150 m. A good overall agreement is seen between the
detector response simulation provided by the two codes.
The apparent differences at large pixel amplitudes result from
very few of the simulated events and are not significant.

\begin{figure}[htbp]
\includefigure{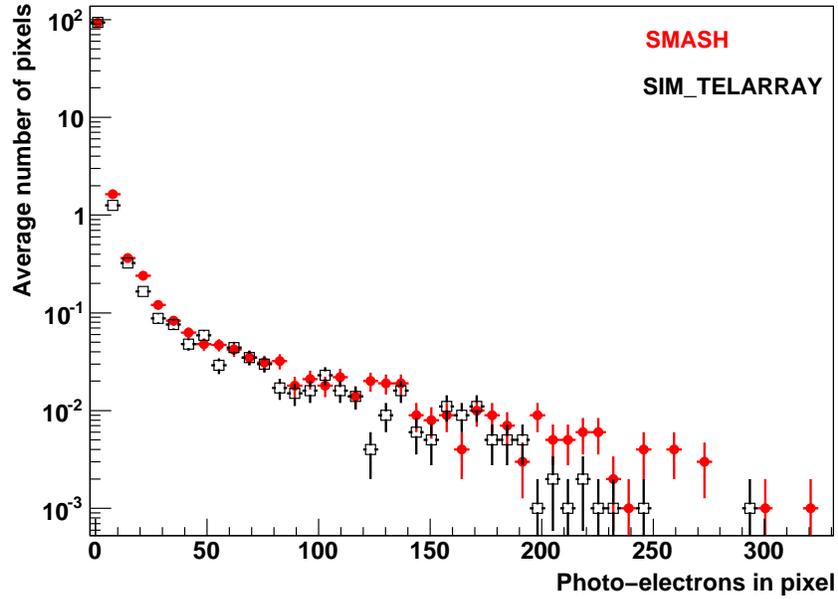}
\caption[Simulated pixel amplitude distribution]
{Average number of pixels with a given raw intensity 
(in units of photo-electrons) for the same
100 simulated air showers seen by \prgname{SMASH} (red filled circles) and
\prgname{sim\_telarray} (black open squares).}
\label{fig:adc_distribution}
\end{figure}


\subsection{Verification of the simulation chain against data}

The simulation chain for CTA is directly based on programs that are
already in use within the H.E.S.S. and MAGIC collaborations and that
have been thoroughly tested  against real data. Comparisons of MC events
against H.E.S.S. data, verifying for example the distribution of photons
in starlight images, the point spread function (PSF) for different
angular distances to the optical axis, or the telescope and system
trigger rate, have been extensively studied as well as the
resulting shower images~\citep[e.g.][]{ber2008,
fun2004, cor2003, HESS-analysis}.

\begin{figure}[htbp]
\includefigure{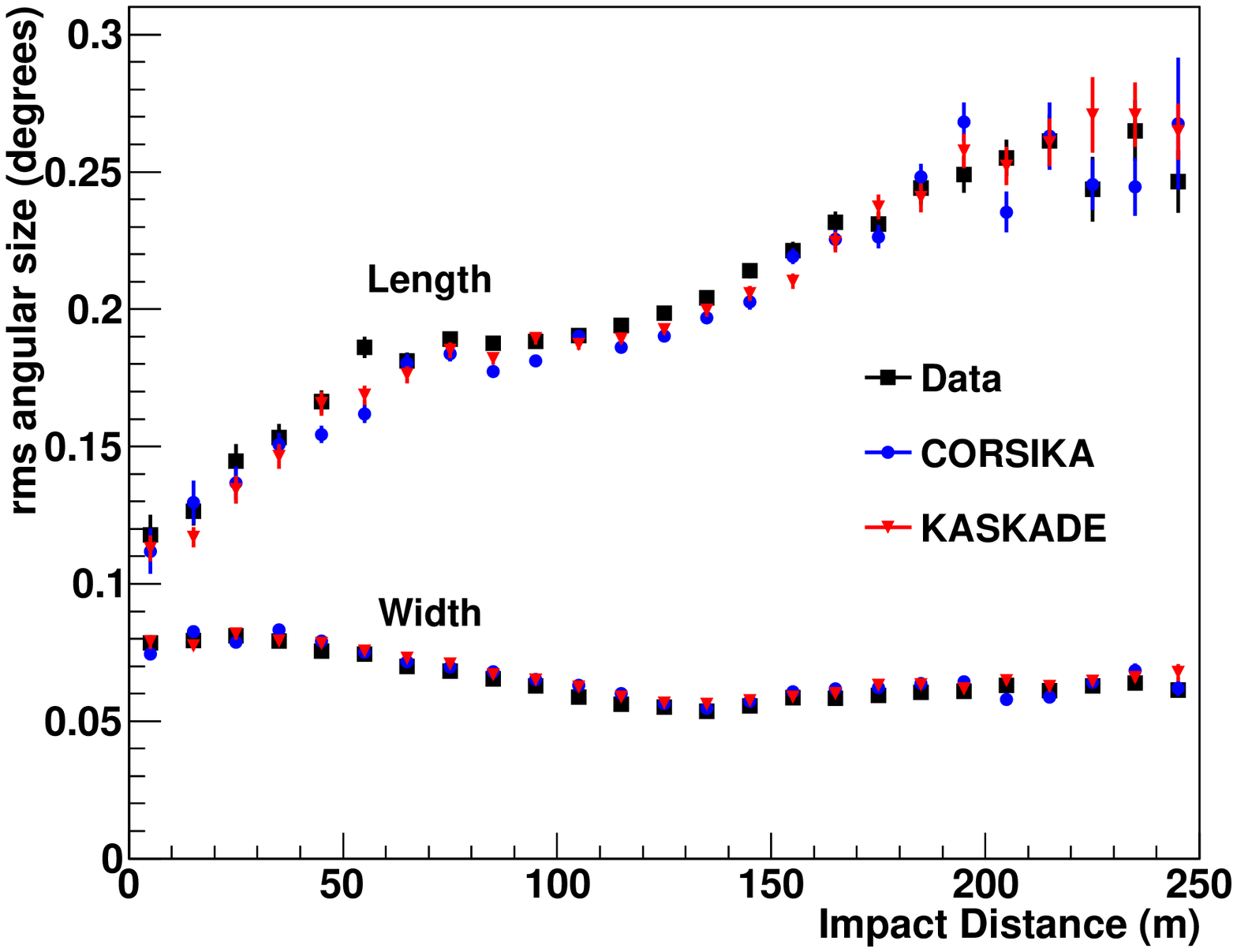}
\caption[Comparison of Hillas parameters]
{Comparison of measured (black squares) and simulated (red triangles and
blue circles) image parameters for the H.E.S.S. telescopes. The real
data are taken from a flare of the blazar PKS~2155-304
\cite{hess_pks2155} for which the signal to noise ratio was very high
and large $\gamma$-ray statistics are available.}
\label{fig:hillas_comp}
\end{figure}

The giant flare detected from the blazar PKS~2155-304 with H.E.S.S. on
July 29th, 2006 provided a good opportunity for an end-to-end test of
the complete simulation chain  for $\gamma$-ray induced showers. The
very high signal to background ratio during the flare, which was
detected at 168 standard deviations in about one and a half hours of
live time, made it appear as an almost pure $\gamma$-ray test beam. For
the data-MC comparison, $\gamma$-ray showers were simulated with the
\prgname{CORSIKA} and \prgname{KASKADE-C$++$} programs and passed 
through the \prgname{SMASH} detector
simulation.  The measured spectrum (power law spectral index) during 
these one and a half hours,
the optical efficiency, the zenith angle
distribution and other runtime parameters were used as inputs to this
simulation to reproduce the exact conditions during  data acquisition.
One of the standard Hillas-type analysis chains was applied to the real
and fake data. Fig.~\ref{fig:hillas_comp} shows the good agreement
(typically at the 5\% level) between the simulated and detected shape of
the shower images, as characterized by their Hillas width and length
parameters. 

\section{Baseline analysis methods}
\label{sec:baseline-analysis}


\subsection { Image cleaning, second moments, and additional parameters }
\label{sec:image-cleaning+Hillas}

The by now quasi classical analysis method for stereoscopic reconstruction
of IACT data is based on the 
{\em Hillas parameters\/} \cite{Hillas}, as derived from 
zeroth order (amplitude or size),
first order (center of gravity position), and 
second order (width, length, orientation) moments
of the images. Since these parameters are highly sensitive to the presence
of NSB noise, {\em image cleaning\/} is applied first, usually in the form
of a two-level (or multi-level) procedure \cite{HEGRA-analysis}. 
The default two-level procedure requires that a pixel is
above a given {\em high\/} level and at least one of its neighbours
is above a {\em low\/} level or vice versa.
The {\em tail-cut levels} of this image cleaning procedure have to be 
adapted to the NSB level, in order to include enough pixels with
significant Cherenkov signals well above the NSB noise level.
Typical high (low) levels are 10.0 (5.0) p.e.\ times
the square root of the per-pixel NSB rate in units of
photo-electrons per 10~ns.

The calibrated and cleaned images are parametrized by the centroid position 
($x_{\rm cog},y_{\rm cog}$) in the camera,
the {\em width} $w$, the {\em length} $\ell$, and 
{\em orientation} $\phi$ parameters of the
Hillas ellipse, as well as the amplitude sum $A$ in units
of p.e. Use of higher-order moments (skewness, kurtosis)
is also possible.

A newer parameter is the {\em time gradient} along the major axis
of the Hillas ellipse. It is obtained from the times when the peak
amplitudes are seen in the individual pixels. This time gradient
is closely related to the distance of the telescope
from the shower axis. It is complemented by pixel time residuals.


\subsection { Geometric shower reconstruction }
\label{sec:basereco-geo}

In the classical Hillas-parameter stereo reconstruction,
the shower direction is determined by a weighted mean of all pair-wise
intersections of the major axes of two suitable images mapped into
a common coordinate system. Suitable here means that images must exceed
a given minimum image amplitude in a minimum number of pixels.
The images should also not be substantially clipped at the 
edge of the camera f.o.v.\ to avoid degraded geometric
and energy reconstructions.

After some experiments, the following
weights for intersection pairs were found to result in a significant
improvement of the angular resolution over previous weighting schemes:
\begin{equation}
w_{ij} = A_{\rm red}^2 \sin^2(\phi_i-\phi_j) \delta_i^2 \delta_j^2,
\end{equation}
where $A_{\rm red}=A_i A_j/(A_i+A_j)$ is the reduced amplitude of the
pair of images and $\delta=1-w/\ell$ is a simplified variant
of the {\em disp} parameter \cite{Fomin-1994a,Lessard-2001}.
The reconstructed shower direction corresponds to the weighted 
average of all intersection points.
The same scheme is also applied to the reconstruction of the shower
core location, which is carried out with telescope positions projected
into the plane perpendicular to the reconstructed shower direction (the 
{\em shower plane\/}).

The selection of images not (much) affected by edge clipping is always
a compromise between the largest possible efficiency and the best possible
angular and core position resolution. The current compromise is:
\begin{equation}
r_{\rm cog} < 0.82 r_{\rm cam} - 0.35 \ell,
\end{equation}
where $r_{\rm cog}$ is the distance between camera center and the image
center-of-gravity, and $r_{\rm cam}$ is the effective radius of the camera,
for the almost circular cameras here the same as their geometrical radius.
There may be room for improvement here, in particular when taking into
account the actual shape of the camera edge. Pixel-based shower fitting schemes
are expected to be affected much less by clipped images than our
baseline scheme.

The {\em height of shower maximum\/} $H_{\rm max}$ or the corresponding
{\em atmospheric depth} $t_{\rm max}$ (the latter measured in g/cm$^2$ 
from the top of the atmosphere
along the shower axis) turns out to be of particular importance for
low-energy showers where the traditional shape cuts for
gamma-hadron discrimination have poor efficiency.


\subsection { Look-up tables for energy reconstruction and gamma-ray selection cuts }
\label{sec:basereco-lookup-tables}

Many of the shower reconstruction and gamma-hadron selection cuts make use of
{\em look-up tables} of the mean values and variances (typically versus two 
independent parameters) of some resulting parameters 
obtained from simulated gamma rays.
These are obtained from filling a total of four corresponding histograms
with (a subset of) the simulated gamma rays, for the number of entries, the
sum of event weights (correcting from simulated spectra to assumed source spectra),
as well as the sum of the event-weighted parameter and its square.

An example is $E/A$, the ratio of (true) shower energy to (measured) image amplitude, 
versus core distance $R_c$ (more precisely the distance of the telescope from the shower axis) 
and $\log_{10} A$ of the telescope. The latter is not
only used to obtain estimates of the primary energy for each telescope with a
suitable image and the expected error on such an estimate, 
but also to obtain a weighted mean energy, to check for the
consistency of the individual estimates and for a possible selection of showers
with high-quality energy estimates. Another example is used for scaling
image widths and lengths for {\em shape cuts} to those expected for gamma rays
(either just scaling to mean 1.0, as used for HEGRA \cite{HEGRA-analysis}, or scaling and
reducing to mean zero and variance 1.0, as used for H.E.S.S.\ \cite{HESS-analysis}),
again versus $R_c$ and $\log_{10} A$.


\subsection { Gamma-hadron selection cuts }
\label{sec:basereco-selection}

The traditional stereoscopic IACT analysis methods use both
image shape and shower direction for discrimination between
gamma-ray initiated showers and hadron showers. The former
({\em shape cuts\/}) are effective for point sources as well
as extended sources but offer no useful discrimination against electron backgrounds.
The latter ({\em direction cuts}) work with any kind of
background but are much more effective for point
sources than for extended sources. 
Our baseline analysis here is optimized for point sources.

A key parameter for further analysis is the number of telescopes 
with cleaned images large and bright enough for further analysis and 
not too close to the camera edge,
$N_{\rm img}$, in the following simply {\em multiplicity}.
The shape cuts use the width and length of each suitable image, 
after converting them to the corresponding {\em reduced scaled width}
and {\em length} \cite{HESS-analysis} 
(assuming an on-axis gamma-ray point-source for filling the
look-up tables used for the conversion).
In contrast to \cite{HESS-analysis}, the {\em mean} reduced scaled width
and length of all $N_{\rm img}$ suitable images in an event are not just
(weighted) mean values but are re-scaled by predefined terms of the form
$\sqrt{a+b N_{\rm img}}$ to achieve a variance close to one 
(and a mean of zero), 
and thus efficiencies of fixed cuts are approximately
independent of multiplicity $N_{\rm img}$ and thus energy $E$.
The actual cuts applied are energy-dependent, because high-energy
hadron showers are easily rejected and larger cut efficiencies are
possible at high energies. At low energies, more strict cuts
will improve $Q=\epsilon_\gamma/\sqrt\epsilon_p$ (the quality
factor for statistical errors)
and $\epsilon_\gamma/\epsilon_p$ (for background systematics),
with $\epsilon_\gamma$ ($\epsilon_p$) being the gamma-ray (proton) cut
efficiency. Helium and heavy nuclei are much more effectively
suppressed by shape cuts and are no longer a significant background
after these cuts. Electron backgrounds cannot be suppressed by
shape cuts. 

The direction cuts for point sources are based on the 
multiplicity-{\hspace{0pt}}dependent
angular resolution, in the form of the 80\% containment radius
$r_{80}$ (to which an additional energy-dependent scaling factor
can be applied but was not used, fixing this factor to 1.0). 
The resulting 80\% cut efficiency
is reasonable but is not optimal over the whole energy range.
In the signal-limited high-energy regime it implies a (not necessary) loss
of 20\% of the signal, while at the background systematics limited
low-energy end, a much tighter cut could improve signal/background
by a modest amount.

Apart from the mandatory shape and direction cuts, there are 
optional cuts, based on the height of the shower maximum ($H_{\rm max}$ cut,
actually cutting on an energy-dependent range in 
the atmospheric depth of the maximum),
on the energy reconstruction quality ($dE$ cut on the
estimated error of the energy) and on the
consistency of energy estimates from the individual images ($dE_2$ cut).
The $dE$ cut takes into account that the energy resolution can be
expected to improve with increasing energy because shower fluctuations
get less relevant, the multiplicity increases and the average
signal in individual telescopes with suitable images increases.  
Events failing the $dE$ cut are typically events with large core
distances to the most nearby telescopes.
Events with inconsistent energy estimates (failing the $dE_2$ cut)
are typically those with bad reconstruction of the core position
and those with a distribution of the Cherenkov light on
ground being very different from the typical shape for gamma-ray showers.

The energy dependence of all adjustable cuts is parametrized as
\begin{equation}
c(E) = \left\{
  \begin{array}{ll}
      c_1, & \textrm{for~} E \le E_1, \\
      c_1 + (c_2-c_1)\frac{\lg E-\lg E_1}{\lg E_2-\lg E_1}, & \textrm{for~} E_1 < E < E_2, \\
      c_2, & \textrm{for~} E \ge E_2.
  \end{array}\right.
\end{equation}
Optimization of the relevant free parameters $c_1$, $c_2$, $E_1$, and $E_2$ 
was performed on an initial
data set of the full configuration with 275 telescopes,
for the given zenith angle, site altitude, NSB brightness
(unless noted otherwise: 20$^\circ$, 2000~m, dark sky,
see Section~\ref{sec:prod1-config} for details). 
No separate optimizations of these
were undertaken for the sub-sets investigated, neither with
the initial nor with the (much larger) corresponding final
simulation data set.

Instead, as a final optimization, there is a choice between
five pre-defined sets of image requirements (amplitude and number of pixels,
dependent on the type of telescope because of different 
per-pixel NSB noise levels
and different pixel scales). The pre-defined $h_{\rm max}$, $dE$, and $dE_2$ cuts
remain optional and are only applied when improving the sensitivity.
The final free parameter is the minimum number of telescopes with suitable
images. These three choices are done separately for each energy interval
and each choice of observation time.
The common picture emerging from this final optimization is:
\begin{itemize}
\item At the lowest energies, very loose cuts are required to get any signal at
all; the low image quality as well as shower fluctuations result in both 
poor gamma-hadron discrimination and poor angular resolution.
Due to the resulting high background,
the sensitivity is typically limited by background systematics,
at least for multi-hour observations and energies close to the detection threshold.
\item At intermediate energies, enough telescopes acquire images of sufficient
amplitude to apply strict image cuts, strict selection cuts, and high
multiplicity. This is the region demanding most CPU time in
simulations, since only a tiny fraction of the hadron showers passes all cuts.
\item At the highest energies, the sensitivity is usually signal limited,
and background does not play an important role. Using as much signal as
possible, with loose cuts and low multiplicity, is of prime importance
for best sensitivity. 
\item Optimal cuts for short observation times tend to be looser than for
long observation times since background systematic uncertainty is less of a problem
and signal limitation sets in at lower energies than for long exposures.
\end{itemize} 


\subsection { Significance of gamma-ray signals above backgrounds }

For the significant detection of a gamma-ray point source the
traditional requirements are a five standard deviation 
(`5-sigma') statistical significance ($S\ge5$)
and the presence of at least ten excess events above background.

For calculating the statistical significance of a gamma-ray
signal above some background, we use equation 17 of Li\&Ma \cite{Li+Ma}.
By convention we assume a signal-free background region 
five times larger than the signal region ($\alpha=0.2$). 
The actual background region in
which we register the -- at high energies very few -- simulated background 
events passing the cuts is independent from that. The registered
(and known to be) signal as well as background events are weighted to 
correct for the different regions as well as for the different spectra in
simulations and nature. The resulting signal and background in the
source regions, with event weights and region corrections applied,
are denoted as $N_\gamma$ and $N_{\rm bg}$ in the following. In the
Li\&Ma notation, $N_{\rm on}=N_\gamma + N_{\rm bg}$ and
$N_{\rm off}=N_{\rm bg}/\alpha$.

%

Due to possible systematics in background subtraction (resulting from
zenith angle, position in the field of view, star light, broken
pixels, or from a non-uniform distribution 
of PMT quantum efficiency or high voltage,
etc.) we also require that the signal excess is at least five times the
assumed background systematic uncertainty of one percent of the remaining
background after cuts. Whether this level of one percent -- better than what
is currently achieved for example with H.E.S.S. \cite{HESS-bg-syst} --
can be reached, may depend on the technical implementation of the
cameras as well as on calibration and monitoring procedures.

In the transition
from statistics-limited to systematics-limited sensitivity, the
baseline analysis as presented here combines the requirements such 
that, for example, a signal of six times the background 
systematic uncertainty and also six times the 
statistical fluctuation cannot result in a detection, despite (barely)
fulfilling the individual requirements.
Apart from this little detail, our requirements for the 
sensitivity limit for any given observation time
(usually 50 hours) can be summarized as
\begin{equation}
\begin{split}
S & \ge  5 \textrm{~(following \cite{Li+Ma} eqn. 17)} \\
N_\gamma & \ge  10 \\
N_\gamma / N_{\rm bg} & \ge  0.05 \\
\alpha & =  0.2
\end{split}
\end{equation}

For the differential sensitivity, we apply these requirements for
each energy interval -- typically with five intervals per decade of
reconstructed energy, i.e.\ intervals of 
the decimal exponent of 0.2 (or 0.2 dex for short) -- 
while for integral sensitivity they are applied
just for the total signal and background. Integral sensitivity
curves as a function of energy are a traditional means for
describing detector performance in VHE $\gamma$-ray astronomy,
but are not used in this paper because they do not show
the actual energy regime where an assumed power-law spectrum
is most significant. Integral sensitivity also heavily
depends on the assumed spectral index and care has to be taken
to not extend it to energies below those where gamma rays
can be readily detected, to avoid extrapolation with the
assumed spectrum. Differential sensitivity does not suffer
from these problems and is therefore preferred although it
depends on the bin size and is not
directly comparable with the integral sensitivity available for
older experiments.

\section{Alternative analysis methods}
\label{sec:alt_methods}

Independent parallel approaches for the improvement in CTA sensitivity 
have been successfully applied to the full CTA MC simulations and are presented in this section.
The standard output of the CTA MC simulation can be processed
with \prgname{read\_hess} \cite{ber2008} and derived programs or it can be converted
into other formats, either at the raw data (ADC counts) level or at the level of
calibrated shower images (pixel-wise charge and arrival time
information). Among the alternate formats are several \prgname{ROOT}-based formats \cite{ROOT} 
which can be further processed for analysis with 
\prgname{MARS}\footnote{The official analysis package of
the MAGIC collaboration \cite{MARS}.}
or \prgname{HAP}\footnote{The official analysis package of the H.E.S.S.\ collaboration.}.
As shown in Sec.~\ref{sec:alt_ana_comparison}, these advanced analysis approaches achieve a sensitivity which is better by a factor of about two with respect 
to the Hillas-based analysis procedure as described in the preceding section.

All analysis methods presented here start by identifying
the pixels with a significant signal with a procedure similar 
to that described in Sec.~\ref{sec:image-cleaning+Hillas}.
The thresholds applied for the cleaning may vary depending on the analysis method.


\subsection{IFAE analysis}

In this analysis\footnote{Developed at IFAE (Institut de Fisica d'Altes Energies)}, 
image cleaning and Hillas parameterization \cite{Hillas} proceed in a 
similar way to the standard analysis, but with somewhat looser criteria 
for inclusion of a pixel in an image (tail-cut levels of 3.0 and 6.0 p.e.), 
and some additional parameters, 
such as the fraction of the total light contained in the two brightest 
pixels ({\em concentration}) also calculated. 
The image amplitude is required to exceed 50 p.e. and the image centroid
is required to be in the inner 80\% of the radius of the field of view.

The shower axis direction and impact point on the ground are estimated 
by a minimisation technique. Energy reconstruction makes use of 
look-up tables in a similar way to the baseline analysis, but with an 
additional dependence on the shower maximum. For background 
suppression this analysis makes use of the multivariate 
classification method known as Random Forest (RF), as used in the 
standard MAGIC analysis \cite{randomforest}. 
To exploit the information from all the 
telescopes triggered in a CTA event, one RF (containing 100 trees) 
for every type of telescope in the simulation is built, using the image 
parameters as input variables ({\em size}, {\em width}, {\em length}, {\em concentration}) 
together with geometric quantities obtained from the stereoscopic 
reconstruction. The output of the RF is a single 
value in the range $[0.:1.]$ for each image, 
which describes the {\em hadronness} of an event.  The global event 
{\em hadronness} is found from the weighted average of the individual 
telescope values. Finally, for every bin in reconstructed energy, 
the cuts in (global) {\em hadronness} and in the squared angular distance 
from the reconstructed event direction to the nominal source position 
are scanned in a range of gamma-ray efficiency from 40\% to 90\%, 
to optimize the point-source flux sensitivity in each bin. 


\subsection{SAM analysis}

The shower axis maximisation (SAM) analysis differs from the 
baseline in that additional information is used for the 
reconstruction of the shower axis and in the methods used 
for background rejection \cite{Parsons-PhD}. 
The method begins with image cleaning and Hillas parameterization 
that are identical to the baseline approach. An iterative approach 
to find the best shower axis (a four-dimensional likelihood 
maximisation) with look-up tables filled from MC gamma-ray 
simulations used to define the expected mean (and probability 
distribution) of image parameters for a given trial shower axis. 
This fit procedure \cite{Hinton} is performed using the following parameters: 
the orientation of the shower images in the camera (similar to 
standard reconstruction), the gradient of pixel trigger times in 
the shower image, the image centroid displacement from the 
trial shower origin and finally the consistency of energy estimates 
between telescopes. The event likelihood (goodness-of-fit) is 
used as an additional gamma/hadron separation parameter. 

Once the event reconstruction has been completed, background 
rejection is performed by use of a neural network (NN), built using 
the \prgname{ROOT} TMVA \cite{TMVA} framework. 
A NN is created for each telescope type, 
using telescope-wise parameters (including the goodness-of-fit), 
and all telescopes within a single event passed through their respective 
network. As for the IFAE analysis, the resulting event classification 
is a weighted average of classifications from all triggered telescopes. 
An energy-dependent cut is then made on this classifier, with the cut 
value being optimised to produce the highest differential sensitivity in 
each energy bin.


\subsection{Paris-MVA analysis}

In the Paris-MVA (multi-variate analysis) approach a 3D-model reconstruction 
\cite{marianne} is coupled to a TMVA background rejection to achieve
 improved sensitivity at low energies as already demonstrated for 
 H.E.S.S.~\cite{Becherini} and then adapted to CTA \cite{BecheriniCTA}.
 Camera images are cleaned and parametrized 
 as described for the baseline analysis (tail-cut levels 5.0 and 7.0 p.e.), 
 with an initial direction estimated from the baseline shower-axis reconstruction. 
 This initial shower axis is refined using the 3D-model reconstruction 
 technique which models the Cherenkov photon emission in the atmosphere 
 by a 3D-photosphere, assumed to have a Gaussian distribution along all axes. 
 This model is used to predict the distribution of Cherenkov light in the 
 cameras of a telescope array by adjusting the intrinsic parameters of the 
 shower to achieve a good fit using a maximum-likelihood optimization of 
 the ensemble of shower images. Energy reconstruction is performed in 
 a similar way to the IFAE analysis
 but the method for the energy evaluation per telescope is somewhat 
 different (see \cite{Becherini} for details).

Background discrimination is performed through an energy-binned 
multivariate procedure using Boosted Decision Trees (BDT) again 
implemented in the TMVA framework. The standard discriminant parameters 
used are the
reduced-scaled Hillas width and length, the reduced 3D-width, its error, 
and the best-fit depth of shower maximum (3D-depth). 
Three new parameters have been defined which identify incoherencies in 
the 3D-model shower fitted images with respect to the observed images. 
These additional parameters concern the angle between the reconstructed 
shower directions for the fitted versus observed images, the
ratio of the energy estimates calculated for fitted versus observed images, 
and finally the deviations between 
measured image size and that expected from the look-up tables.
Best-possible 
signal-to-noise ratio in each energy bin is obtained via a cut made on 
the BDT classifier output and an energy-dependent angular cut.

\section{Interface to the physics working group}
\label{sec:mc2phys}

The physics working group has the goals of preparing the physics program of
CTA, identifying key science programs, and defining benchmark physics targets and
the required instrument performance for them. In order to explore the expected
physics results for selected case studies and the impact of the
instrument performance on the details of the physics output, simulations
of physics cases using realistic CTA performances are required. To enable
these detailed physics studies, the MC working group provides so-called
``CTA performance files'' in ROOT format, which describe the response of a
given CTA layout. A set of tools is then provided to simulate specific
physics cases, such as an energy spectrum of a source, a light curve, or a
spatial morphology of the astrophysics phenomenon.

\subsection{Performance files}

The CTA performance files include histograms to describe in sufficient
detail response functions of a given CTA layout, generally functions of
the reconstructed energy (with five bins per decade) and of the
offset angle with respect to the camera center. In particular, they
include the point source differential sensitivity, the remaining
background rate (for point sources as well as per square degree),
the effective area (both as a function of reconstructed as well as
of true energy), the angular resolution (68\% and 80\% containment),
the energy resolution (r.m.s.), and also the two-dimensional
energy migration matrix.

The response functions depend on the analysis chain used and optimization
criteria applied. Also, the response functions depend on the altitude of
the foreseen observatory, on the night sky background brightness, and 
on the zenith angle of the simulated observations.

\subsection{Tools for simulations of the physics cases}

Several tools have been created to enable a homogeneous comparison between
specific physics cases and individual studies of the required performance.
The tools include:

\begin{itemize} 
   \item  Simulation of an energy spectrum of a user
specified source at a user specified offset from the camera center. The
simulated source might be a point like source or an extended source with
various morphologies. 
   \item  Comparison between two simulated spectra.
This tool aims for a solid statistical comparison between two physics
scenarios in order to answer the question if CTA will be able to
distinguish between them. 
   \item  A sky map tool to represent CTA response
to extended sources of different morphologies and offsets from the camera
center 
\end{itemize}

All tools have the same concept, which is described here. The
user has to provide an energy spectrum of the source, spatial morphology
and the observation time. The photon flux rate is calculated in bins of
offset distance from the camera center and is folded with the
corresponding effective collection area of the CTA array. The energy
migration from true gamma-ray energy into the reconstructed energy is done
according to the migration matrix, which is re-weighted depending on the
spectral shape of the input distribution. The expected number of events in
the signal region is calculated by summing up gamma rays from the simulated
source and expected background level in a particular offset bin. Both 
signal and background event numbers are
randomized according to Poisson statistics. The number of excess
events and its error are calculated by assuming that the background level
is estimated in a five times larger region than the signal region. The
significance of the excess is calculated according to the prescription in
\cite{Li+Ma}. Only bins that fulfil the following criteria are accepted:

\begin{itemize} 
   \item  Significance of the excess in the bin is above 3.0 sigma.
   \item  Number of excess events in the bin is ten or more.
   \item  The excess is larger than 3\% of the background in the bin.
\end{itemize}

The resulting histograms (spectra, integral fluxes, sky maps) can be used
for further analysis and are used to judge the power of a particular
CTA array.
\section{The Production-1 Configuration}
\label{sec:prod1-config}

The goal of the first CTA mass production of simulations was to
characterise the performance of as many, and as varied, CTA candidate
configurations as possible. Memory constraints on the computing nodes
limited the simulation to a total of 275 telescopes, with only a small
fraction of these telescopes being used in any given candidate
array. Most simulations were made for an altitude of 2000~m (typical
of several sites under consideration) and with an geomagnetic field
strength and orientation intermediate between that found in southern
Africa and the Canary Islands. NSB levels usually correspond
to dark sky (a remote site like H.E.S.S. or MAGIC, no moon light,
and a sky region off the Galactic Plane).
Some simulations were also carried out for higher altitude sites
(see Section~\ref{sec:high-altitude}) or for a 
brighter night sky (partial moon light, see Section~\ref{sec:moon-light}).
Most of the analysis work presented
in this paper is based on simulations at $20^\circ$ zenith angle,
while simulation data are also available for $50^\circ$ zenith angle.
These simulations include billions of showers, each used multiple times
at different impact positions, resulting in well over hundred
billion events (mainly protons as the dominating background), with
approximately one out of a thousand events resulting in a stereo
trigger of two telescopes or more.

\begin{table}[tbp]
\caption[Telescope types]{Geometrical parameters of the three main telescope 
types assumed in simulations.}
\begin{tabular}{l|ccc}
 & Large & Medium & Small \\
 & (LST) & (MST$^\dagger$) & (SST) \\
\cline{1-4} \\
 Diameter $D$ (m) & 24.0 & 12.3 & 7.4 \\
 Dish shape$^\ddagger$ & parab. & DC & DC \\
 Mirror area (m$^2$) & 412 & 100 & 37 \\
 Mirror tiles & 594 & 144 & 120 \\
 Tile diam. (m) & 0.90 & 0.90 & 0.60 \\
 Focal length $f$ (m) & 31.2 & 15.6 & 11.2 \\
 $f/D$ & 1.30 & 1.27 & 1.51 \\
 f.o.v. diam. (deg.) & 5 & 8 & 10 \\
 Camera diam. (m) & 2.8 & 2.2 & 2.0 \\
 No. of pixels & 2841 & 1765 & 1417 \\
 Pixel diam. (deg.) & 0.09 & 0.18 & 0.25 \\
 Pixel diam. (mm) & 49 (50$^\star$) & 49 (50$^\star$) & 49 (50$^\star$) \\
  & & &  \\
 \cline{1-4} 
\end{tabular}
\vspace*{6pt}\par\noindent
{\footnotesize Notes: The diameter, $D$, is defined by the outermost mirror edges
while the effective diameter of a circle with the given mirror
area would be smaller.
The mirror area is corrected for inclination. Mirror tile and pixel
diameters are flat-to-flat (all being hexagonal). The camera 
diameter is for the camera body used in ray-tracing.\\
$^\dagger$~The MST-WF is a variant with the same mirror as
an MST, except for $f=16.8$~m, a $10^\circ$ f.o.v. diameter and 
$0.25^\circ$ pixels like an SST.\\
$^\ddagger$~Parabolic or Davies-Cotton (DC).\\
$^\star$~Including a 1~mm gap between pixels. }
\label{tab:tel_types}
\end{table}

Five different types of telescope were used in these
simulations, three types with parameters close to those still under
consideration for CTA, the large-, medium- and small-sized telescopes (LST, MST, SST)
and are described in Table~\ref{tab:tel_types}. Three telescope sizes
are required to achieve the very large energy range of CTA
in a cost-efficient way, with
approximate trigger thresholds for the three components at 20 GeV, 100
GeV and 1 TeV. Davies-Cotton optics were used for the MST and SST
telescope types and parabolic optics for the LSTs. The better off-axis
performance of Davies-Cotton optics is important for the wider f.o.v.\ telescopes, 
whereas the negligible time-spread introduced by the
parabolic optics is more important for the modest f.o.v.\ LST. 
These three telescope types
share a common physical pixel size of 5~cm, the idea being to
simplify the design and construction process through shared
photosensors and other camera components. 
The angles subtended by the
pixels in each of these telescope types are then close to the FWHM of
gamma-ray images at the nominal threshold energy of each type. 
A field of view at the upper edge of the range under consideration was
adopted for each telescope type, so that the full range could be
studied by removing pixels from the analysis at a later point. 
A wide f.o.v.\ version of the MSTs, denoted MST-WF, was considered as an
alternative to SSTs, with MST-like dish but SST-like 0.25$^\circ$ pixels 
and 10$^\circ$ f.o.v. (see Table \ref{tab:tel_types}).

In terms of photosensors, the three main telescope types share
hexagonal pixels with 50~mm spacing, with a bi-alkali type 
quantum efficiency of PMTs having a low afterpulsing ratio.
Readout is assumed at 1.0 gigasamples per second (GS/s) with dual gain
and 12-bit ADCs, similar to H.E.S.S.\ cameras, resulting
in a dynamic range from 0.25~p.e. (electronic noise) to
more than 5000~p.e. (saturation in the low-gain channel).

\begin{figure}[htbp]
\includefigure{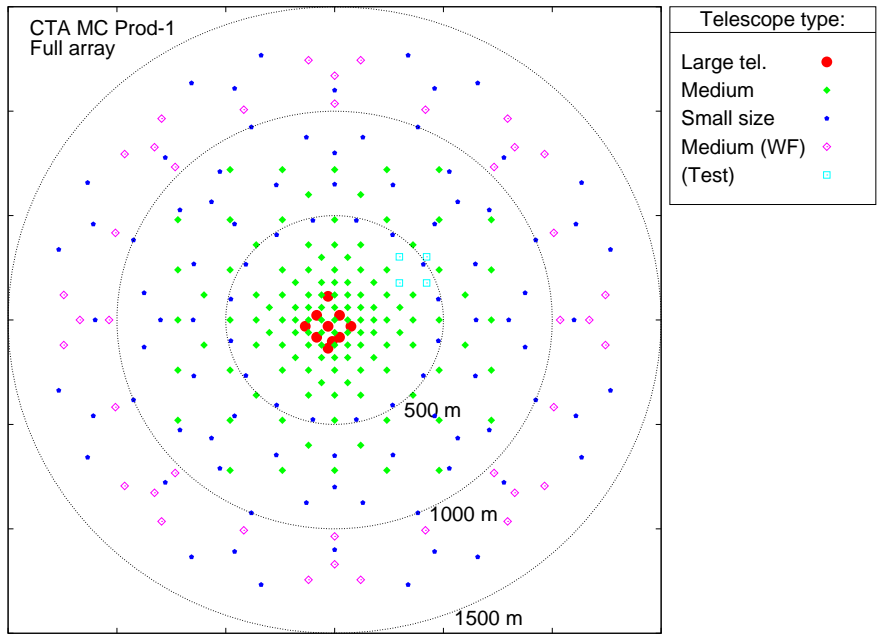}
\caption[Configuration used in simulations]{
The 275-telescope configuration used in the simulations
described here, with LSTs shown in red, MSTs in green and SSTs in blue
(see Table~\ref{tab:tel_types} for details), additional telescopes are
shown in magenta and cyan. The dashed circles
illustrate radii of 0.5, 1.0 and 1.5~km. 
}
\label{fig:prod1layout}
\end{figure}

Figure~\ref{fig:prod1layout} shows the telescope layout chosen for
these simulations. For the low-energy domain of the LSTs, effective
collection area is less critical than a low trigger threshold and
the best possible background rejection power. As a consequence, the LST
component is made up of a small number of large (24~m)
telescopes with moderate separations (less than the Cherenkov shoulder
radius of $\sim120$~m at 2000~m altitude). The final layout
incorporates ten LSTs, allowing subsets of 
two to six telescope
combinations to be selected with different spacings. Possible LST
subsets include both squares and equilateral triangles of both 
75 and 105~m side length.

The MST subarray provides most of the sensitivity in the core energy
range of CTA (0.1-10 TeV) with $\sim25$ telescopes of 12-m class.
For the MSTs the gamma-ray collection area is a critical factor,
and the best trade-off between event-quantity (large area coverage)
and event-quality (high-telescope-multiplicity events) is not obvious
without detailed simulations. As a consequence, a wide range of
spacings for the MST component were tested. The 275 telescope
configuration incorporates regular grids of up to 45 telescopes with 
60, 85, 120, 170, and 240 m spacings and a 25 telescope array with 
340 m spacing, as well as a huge number of multi-baseline alternatives.

The SST telescopes are required to provide a multi-km$^{2}$
collection area above a few TeV. Possible solutions include $\sim$3\,m 
telescopes of moderate ($<$150 m) spacing or more widely spaced
larger telescopes. For classical photomultiplier tube (PMT) cameras and 
telescopes much smaller than 12~m, the
camera cost dominates and very small telescopes are disfavoured
relative to widely spaced telescopes of larger size. The 7.4~m size
adopted is close to optimal in terms of maximum area coverage at a fixed
total array cost. The SST array  is arranged around the MSTs, with
grid-like and island-like layouts incorporated, with 180~m and 240~m spacings.

The candidate array layouts described in Table~\ref{tab:layouts} were 
selected from the production-1 configuration with the goal of exploring a
wide region of the phase space in terms of the balance of sensitivity
across the CTA energy range and the trade-off between event quality
and quantity. CTA South candidates were chosen to have a fixed
nominal telescope construction cost of 80~M\euro{} (in 2005 \euro) and to
have significant sensitivity beyond 100 TeV. Northern array candidates 
NA and NB
have half this nominal cost and no (or almost no) highest-energy component. 
Arrays E and I can
be considered as base-line {\it balanced} layouts, in terms of the
distribution of resources across the full CTA energy range. Arrays A,
B, F and G are more focussed at low energies and C, D and H at high
energies. NB is a higher energy focussed alternative to NA.

\begin{table}[tbp]
\caption[Candidate arrays]{Candidate arrays for CTA South (A-K) and 
  CTA North (NA, NB). For each
  telescope size the number of telescopes of that size is given,
  together with the field of view used in the analysis.}
\begin{tabular}{l|ccc}
   & LST & MST & SST \\ \hline 
A  & 3 ($5^{\circ}$)   & 41 ($8^{\circ}$)   & - \\
B  & 5 ($5^{\circ}$)   & 37 ($8^{\circ}$)   & - \\
C  & -                 & 29 ($8^{\circ}$)   & 26 ($10^{\circ}$)$\dagger$ \\
D  & -                 & 41 ($7.4^{\circ}$) & 16 ($10^{\circ}$)$\dagger$ \\
E  & 4 ($4.6^{\circ}$) & 23 ($8^{\circ}$)   & 32 ($10^{\circ}$) \\ 
F  & 6 ($4.8^{\circ}$) & 29 ($6.3^{\circ}$) & - \\
G  & 6 ($5^{\circ}$)   &  9 ($8^{\circ}$)   & 16 ($10^{\circ}$)   \\
H  & -                 & 25 ($7^{\circ}$)   & 48 ($10^{\circ}$) \\ 
I  & 3 ($4.9^{\circ}$)   & 18 ($8^{\circ}$)   & 56 ($9^{\circ}$) \\
J  & 3 ($4.9^{\circ}$) & 30 ($8^{\circ}$)   & 16 ($9^{\circ}$)$\dagger$ \\
K  & 5 ($5^{\circ}$) & -                  & 72 ($9.5^{\circ}$) \\
NA & 4 ($5^{\circ}$)   & 17 ($6^{\circ}$)   & - \\
NB & 3 ($5^{\circ}$)   & 17 ($6^{\circ}$)   & 8 ($8^{\circ}$) \\ \hline
\multicolumn{4}{l}{\footnotesize{$\dagger$~With wide-field versions of MSTs instead of actual SSTs.}} 
\label{tab:layouts}
\end{tabular}
\end{table}

\section { Performance of different layouts with the baseline analysis }
\label{sec:baseline-performance}


\subsection { Assumptions on source and background spectra }

Simulations were set up to generate primary particles of
power-law differential spectra following $E^{-2}$ (requiring about the
same CPU time per decade in energy for the shower simulations)
or even harder ($E^{-1.7}$, to get enough showers at the highest energies). 
All background particle spectra and most astrophysical
gamma-ray source spectra are substantially softer than $E^{-2}$, and they
may not even follow a power-law. Nevertheless, we assumed power-law
spectra in the following, except for the electron (and positron) background
(see Table~\ref{tab:spectra}).
The latter was described by a log-normal peak of total flux $L$, median energy $E_p$
and width parameter $w$ on top of an $E^{-3.21}$ power-law
spectrum, without any cut-off -- largely consistent with measurements
but rather conservative at TeV energies. 
Our assumed background spectra are based on measurements by 
BESS \cite{bess-p-he-flux}, 
Pamela \cite{pamela-e-flux},
Fermi \cite{fermi-e-flux} and other experiments \cite{cr-flux-wiebel,cr-flux-pdg}.
Gamma-ray source spectra are typically
assumed to follow an $E^{-2.57}$ spectrum (as in the assumed Crab Units spectrum
used as a sensitivity scale) but other spectra are possible.
All results presented in this paper, unless noted otherwise, 
are based on this $E^{-2.57}$ assumption.

\begin{table*}
\caption[Spectral parameters]{Spectral parameters of assumed 
signal and background differential spectra}
\label{tab:spectra}
{\small
\begin{tabular}{lccccc}
Primary & Norm. & Spectral & Log-n. ampl. & location & scale \\
particle & $^\dagger$ & index & $^\dagger$ & [TeV] &  \\
type & $N$ & $k$ & $L$ & $E_p$ & $w$ \\
\hline
$\gamma$ & var.$^{*}$ & $-2.57$  \\
p & 0.096 & $-2.70$ \\
He & 0.0719 & $-2.64$ \\
N (CNO) & 0.0321 & $-2.67$ \\
Si (heavy) & 0.0284 & $-2.66$ \\
Fe & 0.0134 & $-2.63$ \\
$e^-$ (and $e^+$) & $6.85 \times 10^{-5}$ & $-3.21$ & 
  $3.19 \times 10^{-3}$ & 0.107 & 0.776 \\
\hline
\multicolumn{6}{l}{$d F(E)/d E=N\times(E/\textrm{1 TeV})^{-k} + 
   L/(E w \sqrt{2\pi}) \exp({-(ln(E/E_p))^2/2w^2})$} \\
\multicolumn{6}{l}{(see text for more details on log-normal component).} \\
\multicolumn{6}{l}{$^\dagger$ [1/(m$^2$ s sr TeV)]} \\
\multicolumn{6}{l}{$^{*}$ Gamma-ray sources in 
   [1/(m$^2$ s TeV] or C.U. (Crab Units):} \\
\multicolumn{6}{l}{\phantom{$^{*}$}
   1 C.U. = $2.79\cdot10^{-7}$ m$^{-2}$ s$^{-1}$ TeV$^{-1}$ $\times (E/{\rm TeV})^{-2.57}$.}
\end{tabular}
}
\end{table*}


\subsection { Point source sensitivity on-axis }

The sensitivity of more than 50 different subset arrays extracted from the 
simulation data set with the 275 telescope configuration
were evaluated in the way described in Section~\ref{sec:baseline-analysis}.
Most of these arrays were selected to have an installation cost
of 80 M\euro\ in our cost model. These include arrays A--K for the
main (southern) layout candidates, some compact, some more
extended, some without LSTs, some without MSTs, and some without SSTs
but most of them with three types of telescopes.
Other subset arrays include several 
smaller northern site layout
candidates (NA, NB) and a wide range of MST-only arrays with different
separations and with different fields of view.

Image cleaning (adjusted to the different NSB levels in the different
telescope types), five image selection rule sets (`extra-soft' to `hard',
with different requirements on image amplitude and number of pixels included), 
and the energy dependence of various shower-selection cut parameters
were defined in advance. Integral and differential sensitivities were
evaluated for each of the image selection rule sets, for each multiplicity
of telescopes with useful images from two up to eight, and including
or ignoring the optional $H_{\rm max}$, $dE$, and $dE_2$ selection cuts,
allowing for a final optimization as the last step.

\begin{figure}[htbp]
 \includefigure{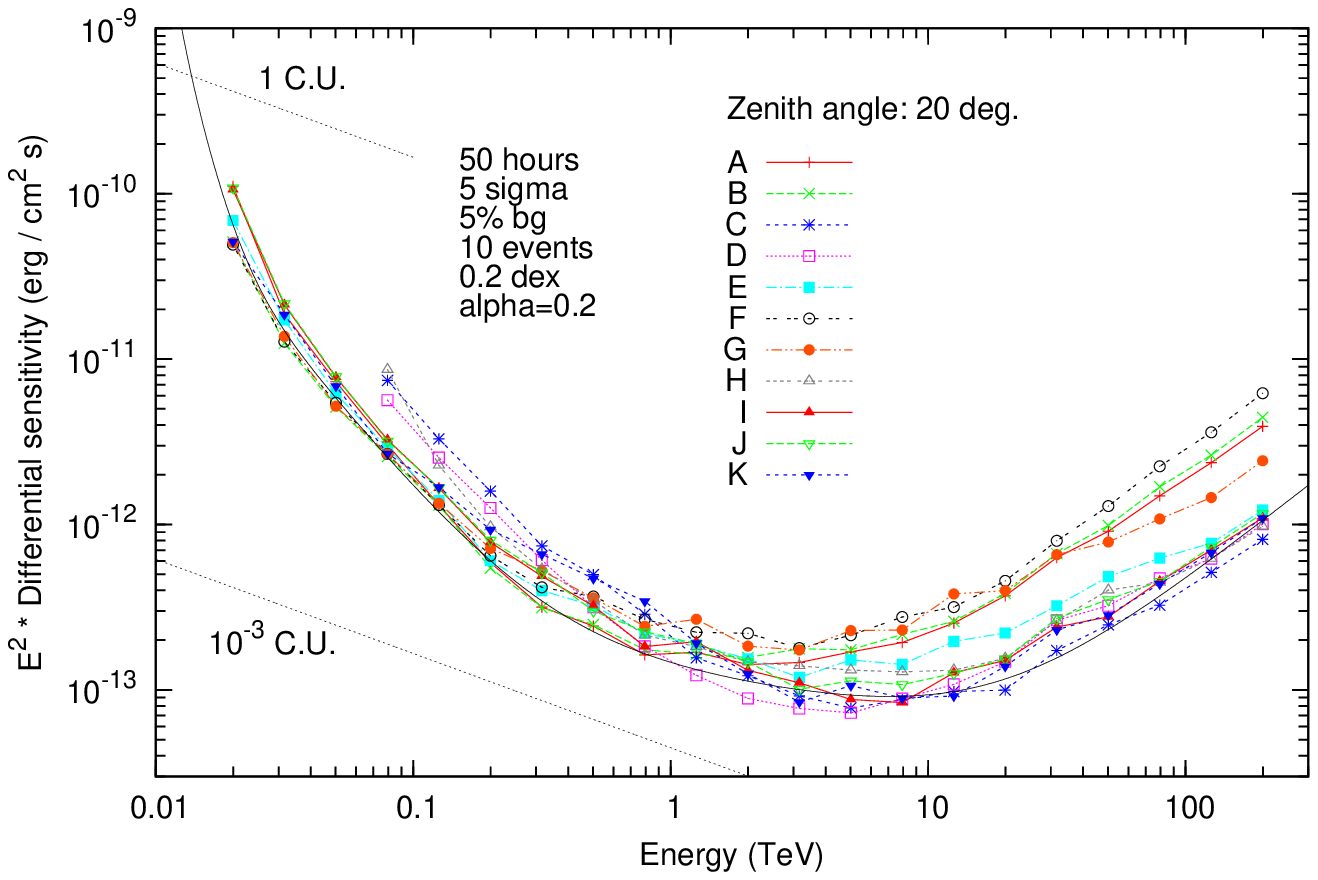} 
 \includefigure{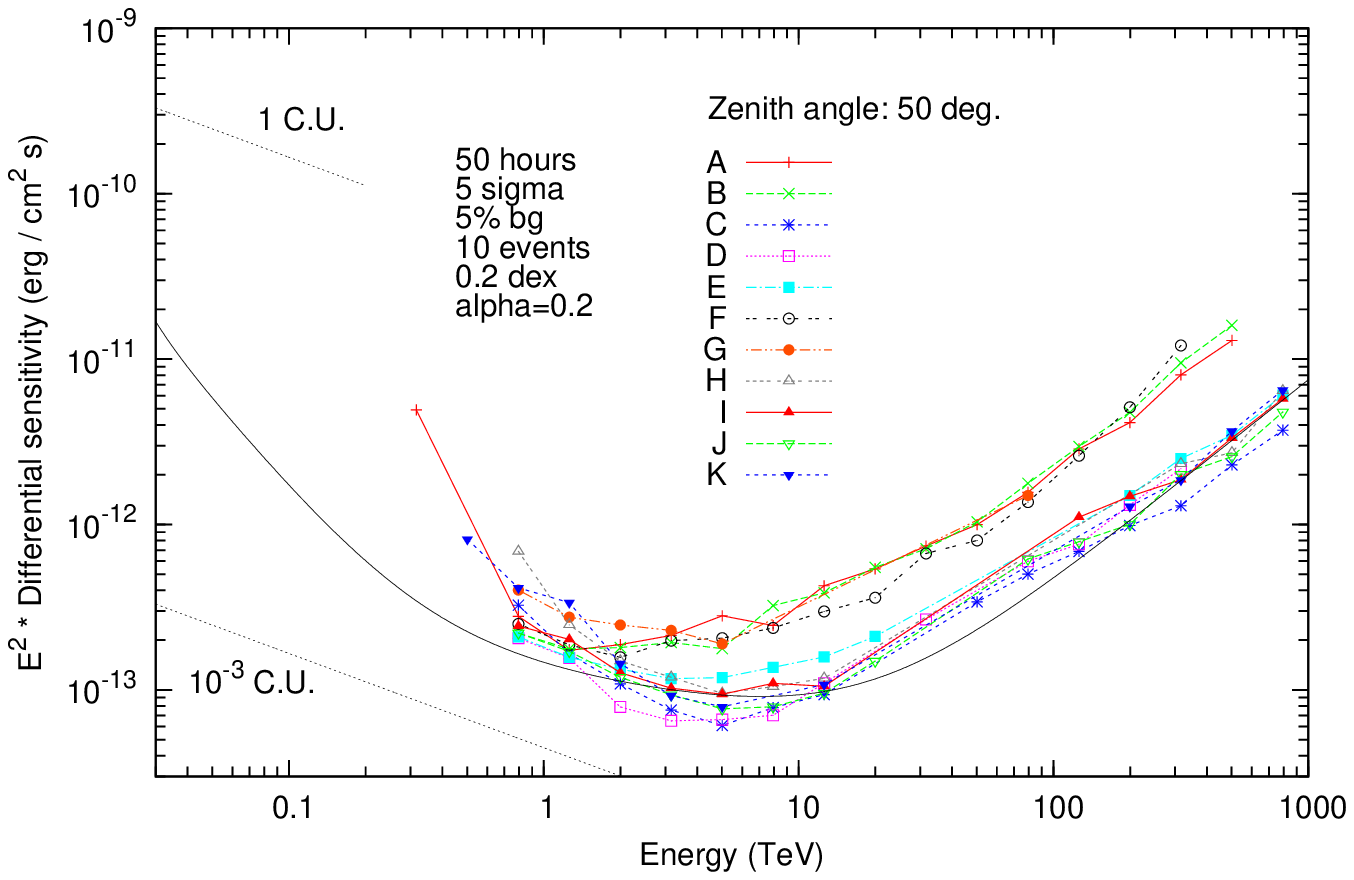}
 \caption[Sensitivity of 11 candidate array layouts]{
   Point source sensitivity of 11 CTA candidate array layouts
   (of identical estimated costs, for CTA South) for
   50 hours observation time, evaluated with the baseline analysis method.
   Note that this differential sensitivity corresponds to an
   independent detection in each energy interval and is much more
   strict than the conventional integral sensitivity -- but almost
   independent of the assumed spectrum.
   The solid black line is an approximation of the best performance
   of any of these arrays at any energy (except {\em D\/}
   which is highly specialized for energies of a few TeV),
   for $20^\circ$ zenith angle.
   The Crab Unit (C.U.) and milli-C.U. fluxes as used in this
   paper are indicated for comparison.
   Top: $20^\circ$ zenith angle, bottom: $50^\circ$ zenith angle.}
 \label{fig:sens_A-K}
\end{figure}

Figure \ref{fig:sens_A-K} shows the on-axis point-source differential
sensitivity of arrays A to K at 20$^\circ$ zenith angle. 
It can be seen that there are
three categories: compact arrays (usually without small telescopes),
extended arrays (without large telescopes), and {\em balanced arrays\/}
which try to find a compromise between the compact and extended cases.
At the lowest energies, the sensitivity curves split up by the number
of large telescopes (5, 4, 3, or none). At the highest energies,
the sensitivity is always signal limited and thus dominated by the
area covered. 

Statistical errors on the derived sensitivity (mainly from the
finite number of proton showers passing cuts) are neither included in 
Figure~\ref{fig:sens_A-K} nor the following since they are highly correlated 
between different array layouts (sharing the same simulated showers and
also part of the telescope data). Statistical errors are actually
smaller than the fluctuations seen from energy bin to energy bin,
which also result from the optimization process (e.g.\ integer
values of multiplicity).

\begin{figure}[htbp]
 \includefigure{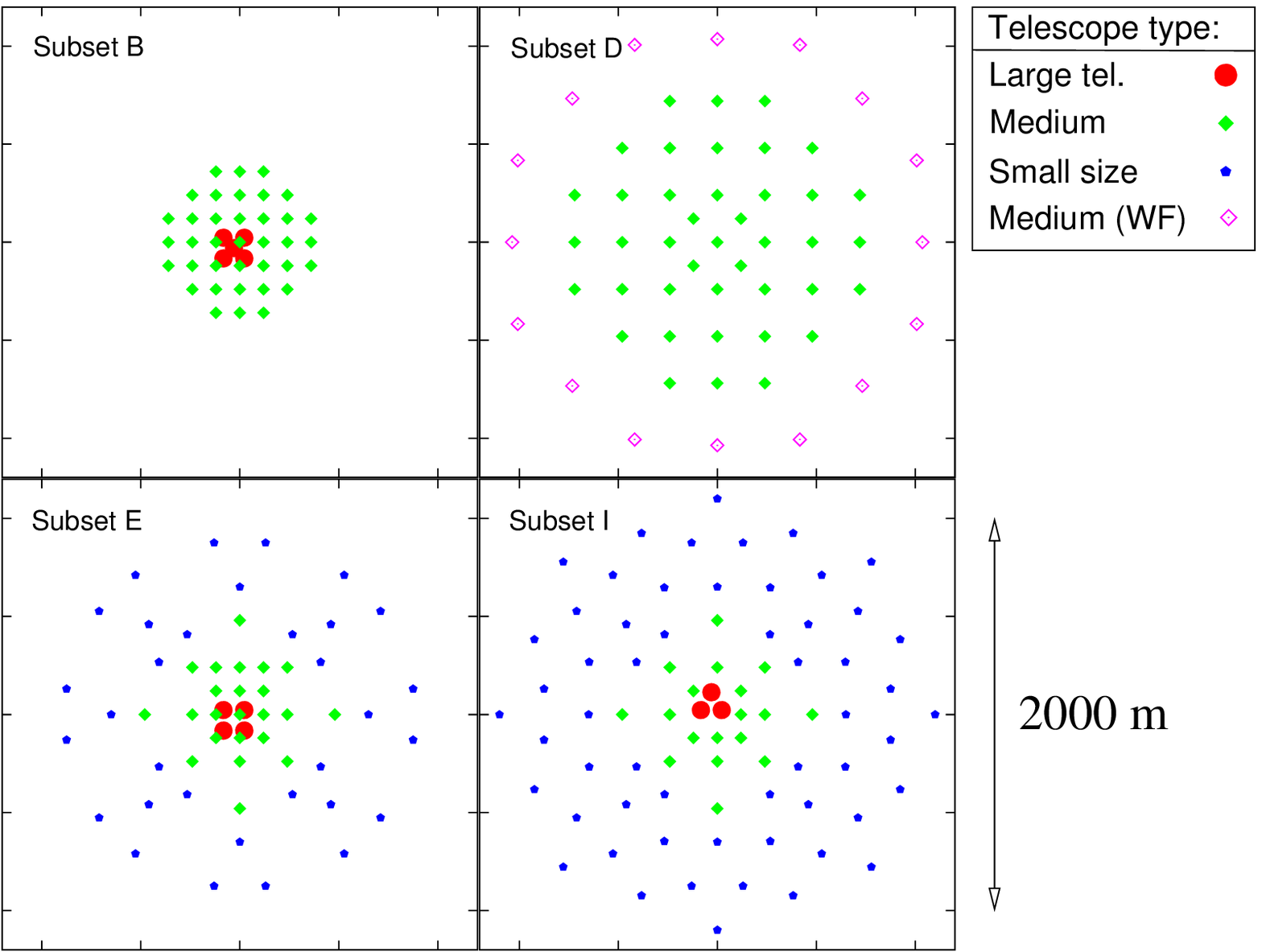}
 \caption[Layout examples]{Layout examples for a compact array layout (B), an extended
  layout without LSTs (D), as well as two balanced layouts (E and I).}
 \label{fig:layouts-selected}
\end{figure}

Good examples of balanced arrays can be seen with
arrays E and I. Array E, with four large telescopes, performs
slightly better at low energies than array I, with only three large 
telescopes. 
At large energies, array I with its extended set of small telescopes
outperforms array E by typically a factor of 1.5. Array B is a typical
case for a compact layout while array D is one of the extended layouts
without any large telescopes (see Figure \ref{fig:layouts-selected}).

The candidate layouts for a northern CTA site, without extended sets
of small telescopes and with fewer mid-size telescopes, are rather 
similar in their performance to the compact full layouts but with 
slightly inferior sensitivity at a TeV and above.

\begin{figure}[htbp]
 \includefigure{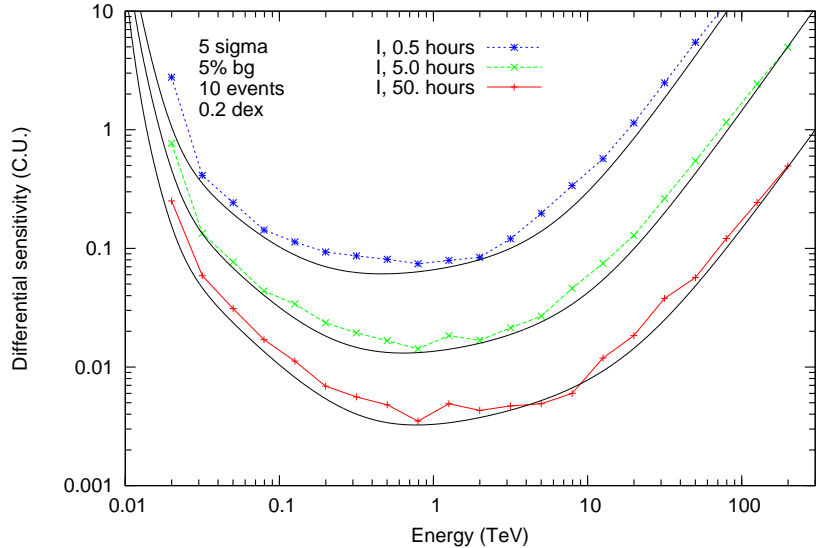}
 \caption[Sensitivity of array I for 0.5 to 50 hours]{Point source sensitivity of array I 
 (in units of 1 C.U. = $2.79\times10^{-7} (E/\textrm{TeV})^{-2.57} 
 \textrm{~m}^{-2} \textrm{~s}^{-1} \textrm{~TeV}^{-1}$) for
 observation times of 0.5 hours, 5 hours, and 50 hours, respectively.
 Also shown as black solid lines are approximations to the best
 performance of any of the 11 CTA South arrays at any energy (as in
 figure \ref{fig:sens_A-K}), for the
 given observation times. Array I, being close to this optimum
 at all energies, is indeed a well-balanced array.}
 \label{fig:sens_obs_time_I}
\end{figure}

The different limiting factors for the sensitivity -- signal, statistics,
or background systematics -- also result in different dependence of the
sensitivity on observation times $T$: proportional to $1/T$ at the highest
energies, $\propto 1/\sqrt{T}$ at intermediate energies, and substantially
weaker than $1/\sqrt{T}$ at the lowest energies 
(see Figure~\ref{fig:sens_obs_time_I}).
Note that the final optimization of image selection, multiplicity,
and choice of optional cuts has been done separately for each
observation time, typically resulting in looser selections for shorter
observation times. Figure \ref{fig:sens_obs_time_I} also demonstrates
that array I is in fact a well-balanced array, with little
sensitivity loss against specialized arrays in any energy range.

\begin{figure}[htbp]
 \includefigure{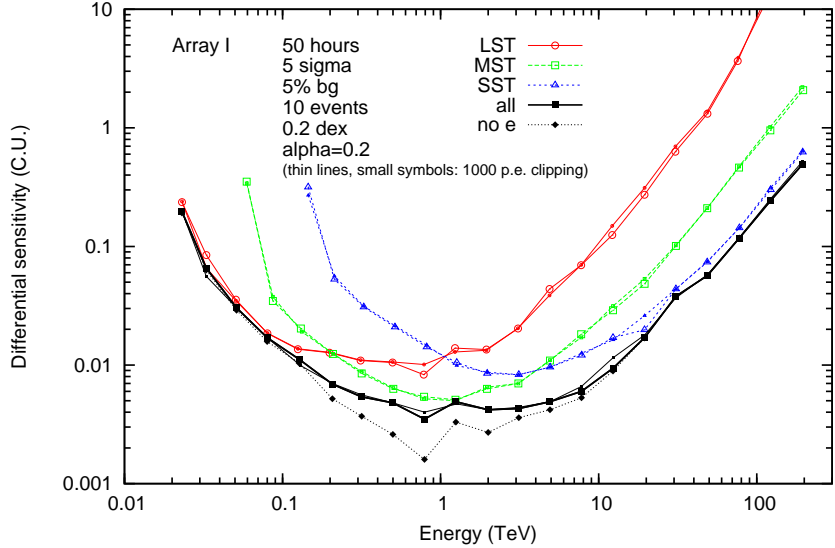}
 \caption[Sensitivity of array I and components]{
 Point source sensitivity of array I (solid black line, 
 filled squares) and
 its components, 3 LSTs (red, open circles), 18 MSTs (green, open squares), 
 56 SSTs (blue, open triangles). Thin lines with small symbols illustrate
 the limited impact of a reduced dynamic range of PMT readout electronics.
 For the relevance of the electron background on the combined sensitivity 
 see also the dashed black line
 with diamonds, where this background is ignored. }
 \label{fig:array_I_components}
\end{figure}

The contribution of the different types of telescopes to the overall
sensitivity for on-axis point sources is demonstrated for array I in 
Figure \ref{fig:array_I_components},
for an observation time of 50 hours.
The cross-over between LSTs and MSTs is seen at about 250 GeV, that
between MSTs and SSTs at about 4 TeV. At these cross-over points, the
larger telescopes contribute fewer but higher quality data while
the smaller telescopes provide a larger effective area but with
lower image quality. At both transition points, 
the combined sensitivity is
almost a factor of two better than that of the individual components.
Near 1 TeV, though, the onset of low-quality SST data, with large
effective area but poor gamma-hadron rejection deteriorates the
combined sensitivity in our simple analysis method basically to the
sensitivity of the MST data alone.

Because no small telescopes were assumed in the inner region of
array I, the MSTs continue to contribute to the combined sensitivity
up to the highest energies, despite possible signal saturation.
At a smaller (10-15\%) level, the presence of MSTs even improves the 
sensitivity at the lowest energies -- where the MSTs are not expected to be
triggered at all by gamma rays or electrons but still can reject some
hadron background. Such background includes muons seen by MSTs or 
events where the LSTs only registered a small gamma-like sub-shower of a
larger hadron shower.

The effect of a signal limitation at 1000 p.e.\ per pixel is also
indicated in Figure~\ref{fig:array_I_components} by the thin lines
with small symbols. Even though LSTs above
several TeV and MSTs above a few tens of TeV suffer from pixel saturation,
with an impact on angular and energy resolution (see 
Section~\ref{sec:ang+e_res}), this only happens in a regime where the
sensitivity is signal limited and does not depend on
angular resolution.

Finally, Figure~\ref{fig:array_I_components} also illustrates the relevance 
of the electron background on the combined sensitivity. This background
is most relevant in the region of the `shoulder' in the cosmic ray
electron (and positron) spectrum at a few hundred GeV. Below about 200 GeV,
the rejection of proton-induced showers deteriorates substantially
and the background is dominated by the protons. Above a few TeV,
both the electron and the hadron backgrounds are at a very low level
and the point source sensitivity is signal limited.

The second CTA site, in the northern hemisphere, is foreseen
to be instrumented similar to array I excluding its extended SST component.
As such, a northern CTA site could perform similar to the
main site at low energies (unless affected by the geomagnetic field)
to a few TeV. At energies above a few TeV, such a northern site will
suffer from its much lower effective area.


\subsection { Angular resolution and energy resolution }
\label{sec:ang+e_res}

Angular and energy resolution in an installation with different
instrument types (and separations between instruments increasing
outwards) combine in ways which are not always intuitive.
In a homogeneous array, increasing energy will result
in more usable telescopes with, on average, improved data quality
and therefore improving angular and energy resolution,
at least to the point where the instruments start running into
pixel saturation.

In a realistic array, showers of increasing energy will trigger
smaller instruments in wider separations to their next neighbours.
Most of its effective area may be from showers recorded with
few and widely separated telescopes. A subset of the showers,
typically those with shower cores closer to the array centre,
will be recorded with higher telescope multiplicity and data
quality, providing a subset of high angular resolution and/or
of high energy resolution. Figures \ref{fig:ang_res_components}
and \ref{fig:e_res_components} show only the
angular resolution and energy resolution achieved with
cuts separately optimizing these resolutions, at 
minimum multiplicities between two and six.

\begin{figure}[htbp]
 \includefigure{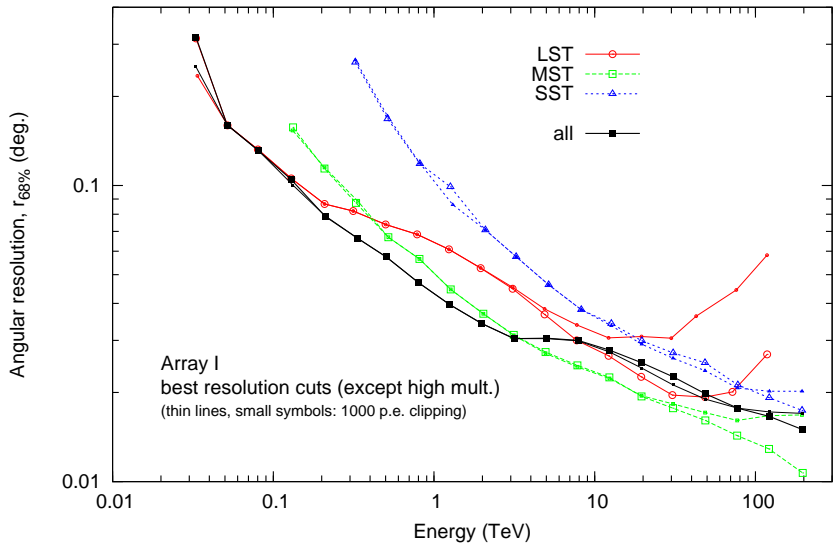}
 \caption[Angular resolution of array I]{
 	Angular resolution of array I (68\% containment radius)
    as a function of energy (lines and symbols as in 
    Figure~\ref{fig:array_I_components}). Cuts were optimized for
    angular resolution, at a minimum multiplicity between two and six.
    At energies above 5 TeV a pixel dynamic range limited to 
    1000 p.e.\ would have substantial impact on the LST angular
    resolution but basically no impact on the resolution of the
    full array.
    While the angular resolution of all telescopes is
    always better than that of the component dominating the
    effective area at a given energy, selecting data from
    a single component may improve angular resolution at
    the cost of a much reduced effective area.}
 \label{fig:ang_res_components}
\end{figure}

\begin{figure}[htbp]
 \includefigure{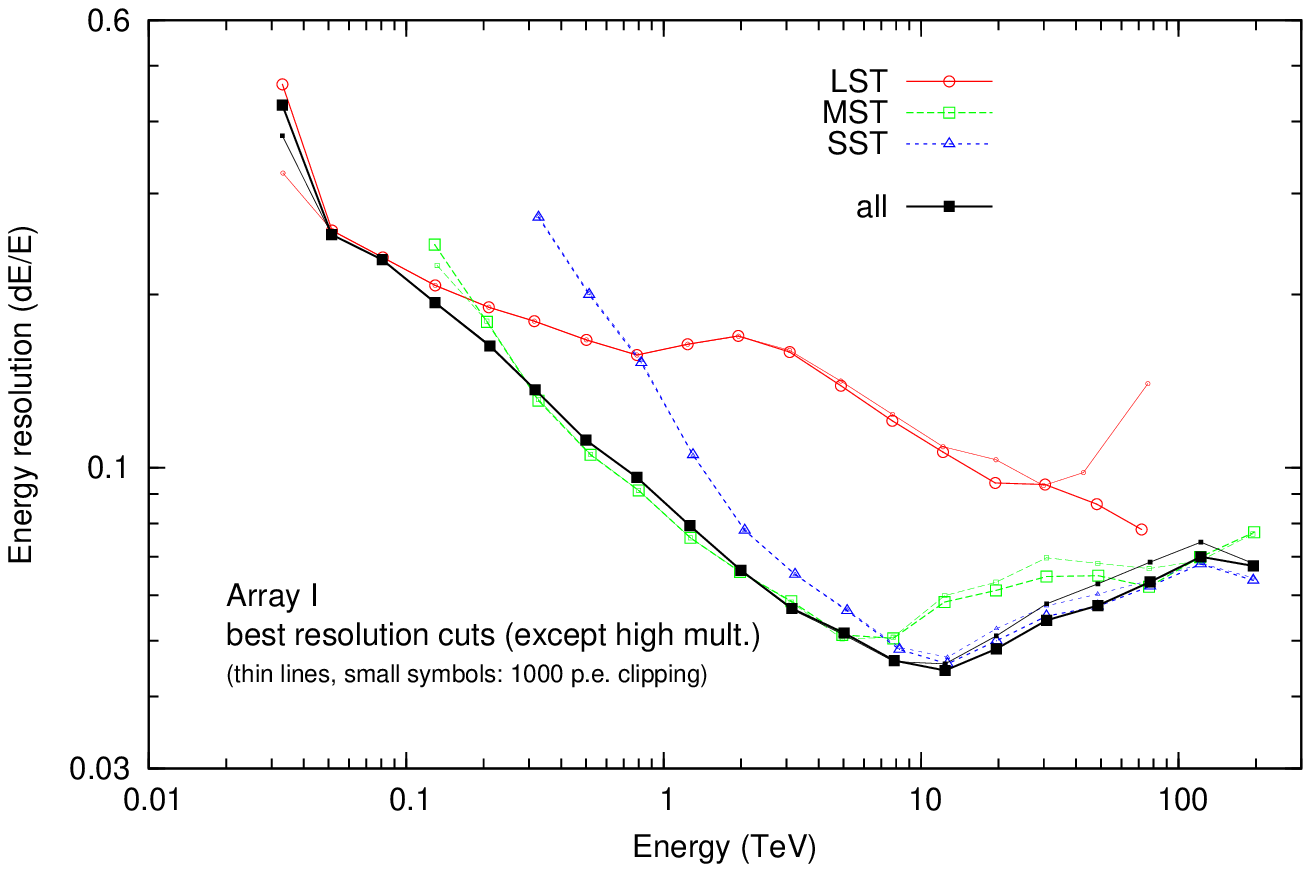}
 \caption[Energy resolution of array I]{
 	Relative r.m.s.\ energy resolution $\sigma(E)/E$ of array I
    as a function of energy (lines and symbols as in 
    Figure~\ref{fig:array_I_components}). Cuts were optimized for
    energy resolution, at a minimum multiplicity between two and six.}
 \label{fig:e_res_components}
\end{figure}


\subsection { MST inter-telescope separation and field of view }

One important question related to the CTA layout is the
correlation between optimum inter-telescope separation and
the f.o.v. of a telescope. The available data allowed to
study this with MST-only arrays of similar cost estimates.
We extracted data for arrays of telescopes at a spacing of either
60~m, 85~m, 120~m, 170~m, or 240~m. At each of these spacings
we used arrays of 37 telescopes with a f.o.v. diameter of 5.0 degrees,
of 32 telescopes of 6.0 degrees, 27 telescopes of 7.0 degrees, as well
as 24 telescopes of 8.0 degrees. Figure~\ref{fig:sep-fov} shows the
resulting on-axis differential point-source sensitivity for 120~m separation.
Except at the highest energies, a f.o.v. of only 5 degrees is typically
best -- but the performance of fewer telescopes with larger f.o.v. is
quite similar. For extended sources or when multiple sources can be
studied at the same time -- the typical case along the Galactic Plane --
the larger f.o.v. comes with additional benefits. For this reason, 
the MST telescopes for CTA are foreseen to have cameras with a f.o.v.
between 6 and 8 degrees.

\begin{figure}[htbp]
 \includefigure{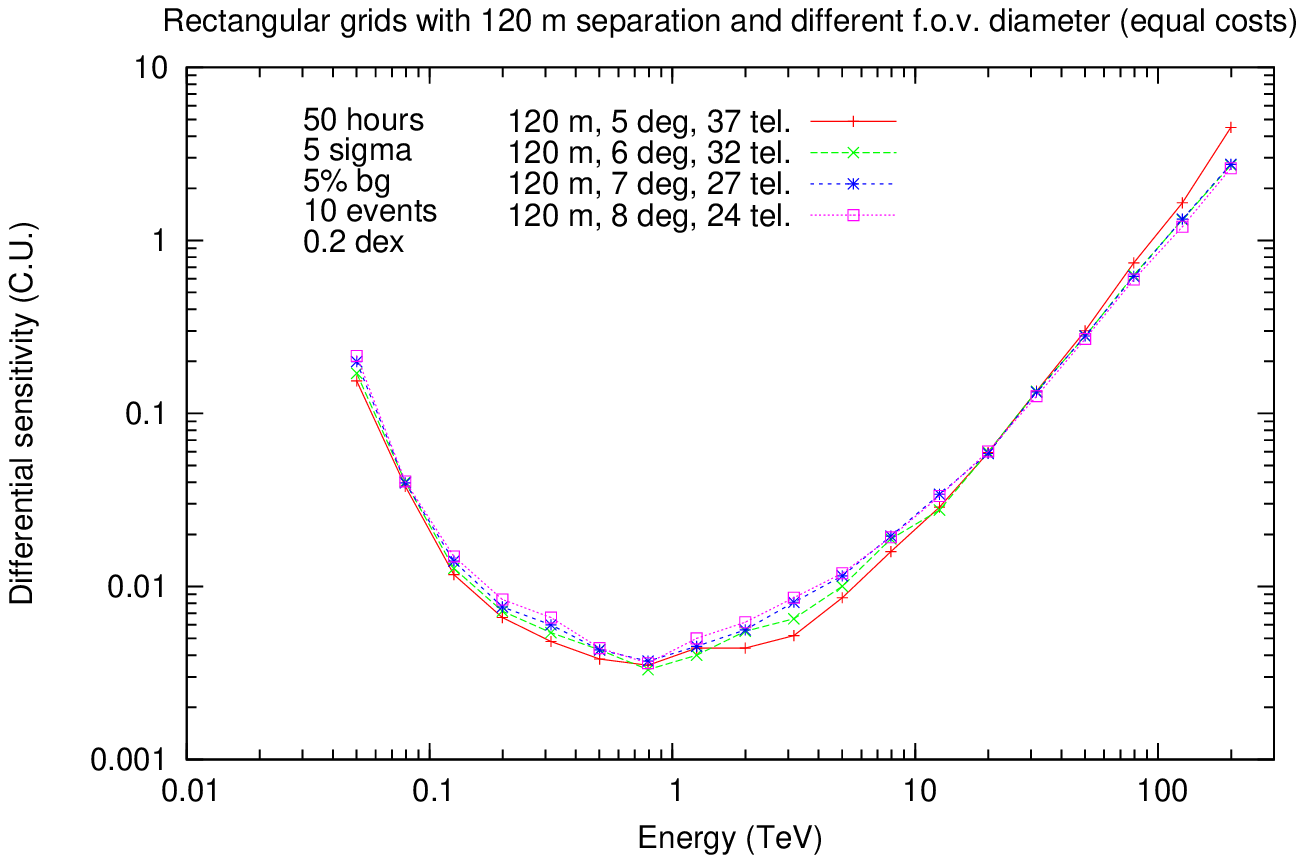}
 \caption[MST array sensitivity with different Fov]{
 	Differential point-source sensitivity on-axis in C.U. for MST
    arrays of similar cost but with cameras of different f.o.v. (fewer
    telescopes for larger f.o.v., see text
    for details), for a 120~m inter-telescope separation and 20 degrees
    zenith angle.}
 \label{fig:sep-fov}
\end{figure}

\begin{figure}[htbp]
 \includefigure{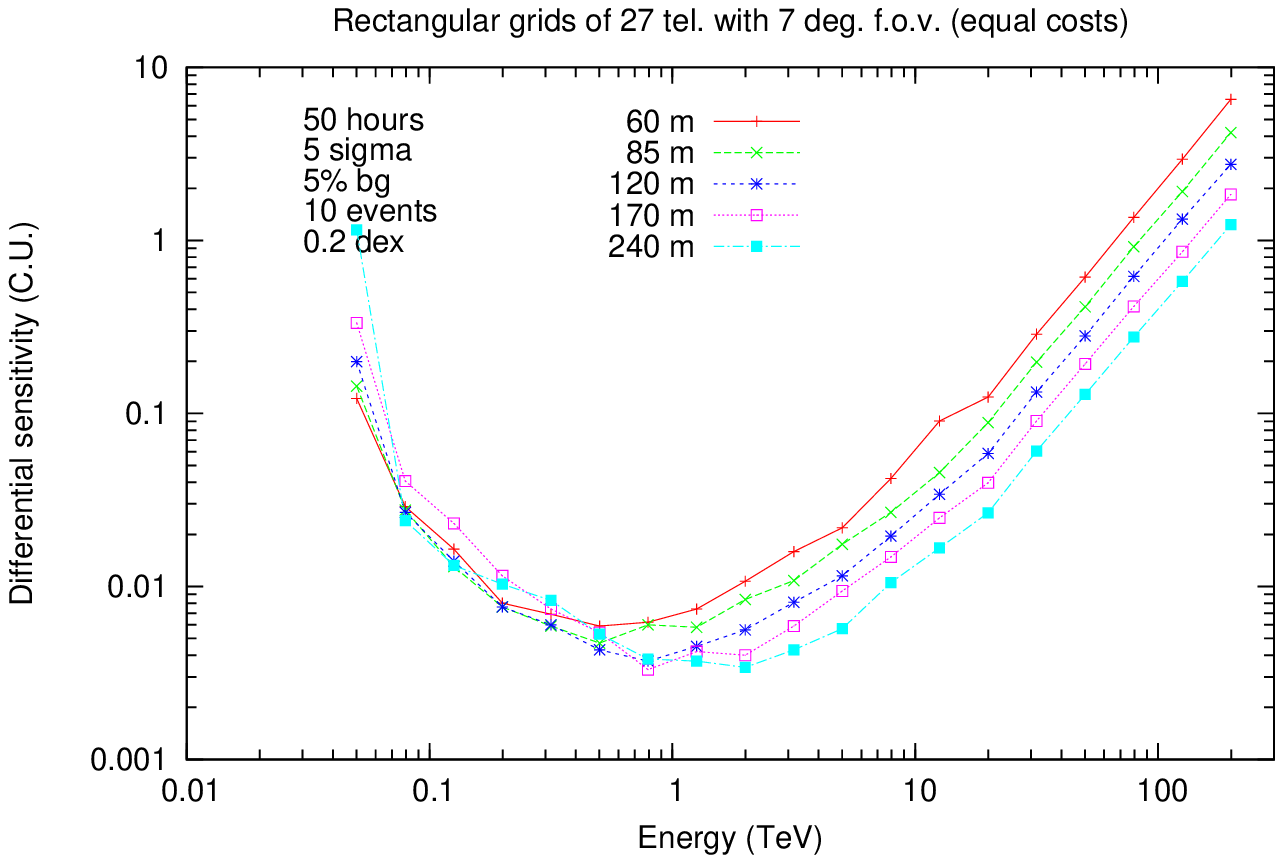}
 \caption[MST array sensitivity with different separations]{
 	Differential point-source sensitivity on-axis in C.U. for MST
    arrays of 27 telescopes with 7 degrees f.o.v., 
    at different inter-telescope separations (zenith angle: 20 degrees).}
 \label{fig:sep-diff}
\end{figure}

In Figure~\ref{fig:sep-diff} we have 7.0 degree f.o.v.\ cameras in 27 telescopes
at different separations. It is obvious that separations below 120~m
have no advantages 
at any energies, except at the very threshold. At energies beyond a TeV, the
larger effective areas resulting from larger separations more than
compensate for the poor sampling of each shower (seen in fewer telescopes).
Since there is no separation that can optimize the performance
simultaneously at all energies, a {\em graded layout} with inter-telescope
separations increasing from the array center outwards will result in
a better overall performance than a regular grid.
Note that at larger zenith angles the Cherenkov light pool on the ground
will increase and optimum spacings are always larger than at small zenith
angles.

\section{Comparisons for candidate array I with alternative analyses}
\label{sec:alt_ana_comparison}

We discuss the expected performance of the candidate array I
that is obtained by using the alternative analyses described in 
Section~\ref{sec:alt_methods}.
The array consists of 3 LSTs with a field of view (f.o.v.) diameter of 4.9$^{\circ}$,
18 MSTs with a f.o.v. diameter of 8$^{\circ}$ and 56 SSTs with a f.o.v. diameter of 9$^{\circ}$,
whose positions on the ground are shown in Figure \ref{fig:layouts-selected}.
We also mention briefly comparisons between different site altitudes
and for observations under partial moon light.


\subsection{The differential flux sensitivity}

The minimum detectable flux is determined, by demanding a minimum $5\sigma$ detection (using Equation 17 from  Li \& Ma \cite{Li+Ma}),
at least ten gamma-ray events, and a gamma-ray excess of at least $5\%$ of the residual cosmic-ray background.
The differential flux sensitivities achieved from the alternative analyses are shown in  Figure \ref{fig7:diffsens},
using five bins per decade in energy.
Differences between sensitivity curves may be understood in terms of the respective effective areas and residual cosmic-ray background rates,
which are shown in Figures \ref{fig7:aeff} and \ref{fig7:backgnd}, respectively.
These effective areas and background rates are in turn dependent on the specifics of each analysis (see Sections~\ref{sec:baseline-analysis} and \ref{sec:alt_methods} for details).
This may include the image cleaning, quality cuts, shower reconstruction, and cosmic-ray background rejection power,
along with the $\gamma$-ray selection-cut optimization scheme that is employed for each analysis.
The optimization is on sensitivity and good sensitivity can be achieved either
with large effective area or low background. Therefore small fluctuations in simulated data can
result in large apparent fluctuations in Figures \ref{fig7:aeff} and \ref{fig7:backgnd}.
The most sensitive analyses (SAM and Paris-MVA) approach levels of $2$ milli-C.U.
in differential sensitivity per bin at energies of around $1$ TeV,
or about one milli-Crab in integral sensitivity.

\begin{figure}
    \includefigure{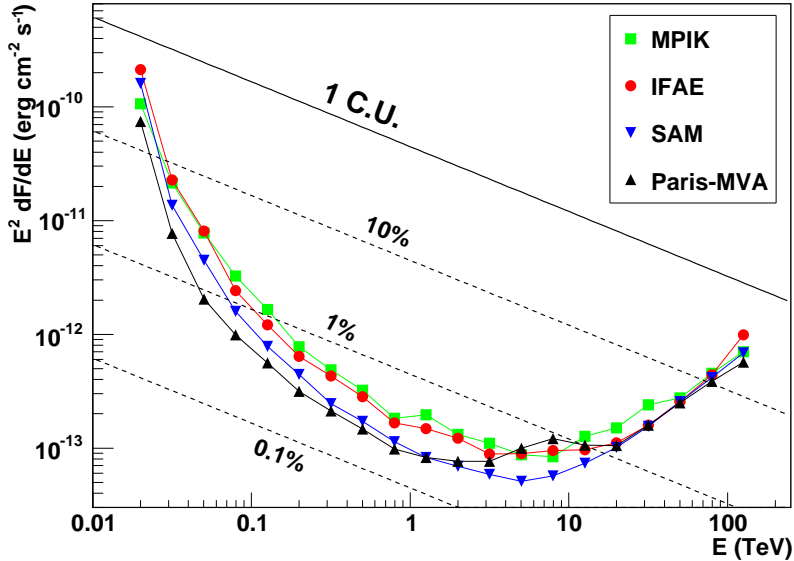}
    \caption[Differential sensitivity in alternate analyses]{
      Differential flux sensitivity of candidate array I
      given as a function of the estimated energy,
      for the baseline/MPIK (green squares), IFAE (red circles),
      SAM (blue triangles) and Paris-MVA (black triangles) analyses.
      The Crab Unit (C.U.) flux (solid black line) 
       is shown for comparison,
      together with its $10\%$, $1\%$ and $0.1\%$ flux levels (black dashed lines).
      The differential sensitivities are optimized for an observation time of 50 hours.
    }
    \label{fig7:diffsens}
\end{figure}

\begin{figure}
    \includefigure{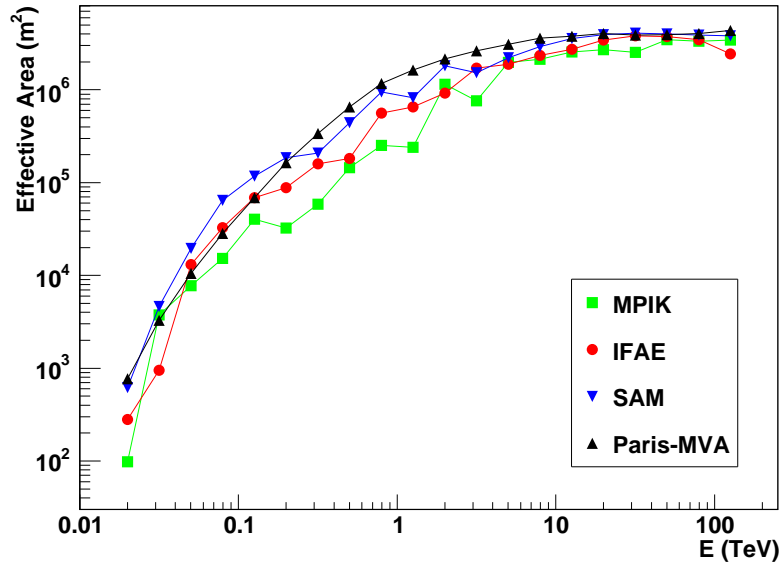}
    \caption[Effective area in alternate analyses]{
      Effective area of candidate array I,
      given as a function of the estimated energy
      for the baseline/MPIK (green squares), IFAE (red circles),
      SAM (blue triangles) and Paris-MVA (black triangles) analyses.
      The differential sensitivities are optimized for an observation time of 50 hours.
    }
    \label{fig7:aeff}
\end{figure}

\begin{figure}
    \includefigure{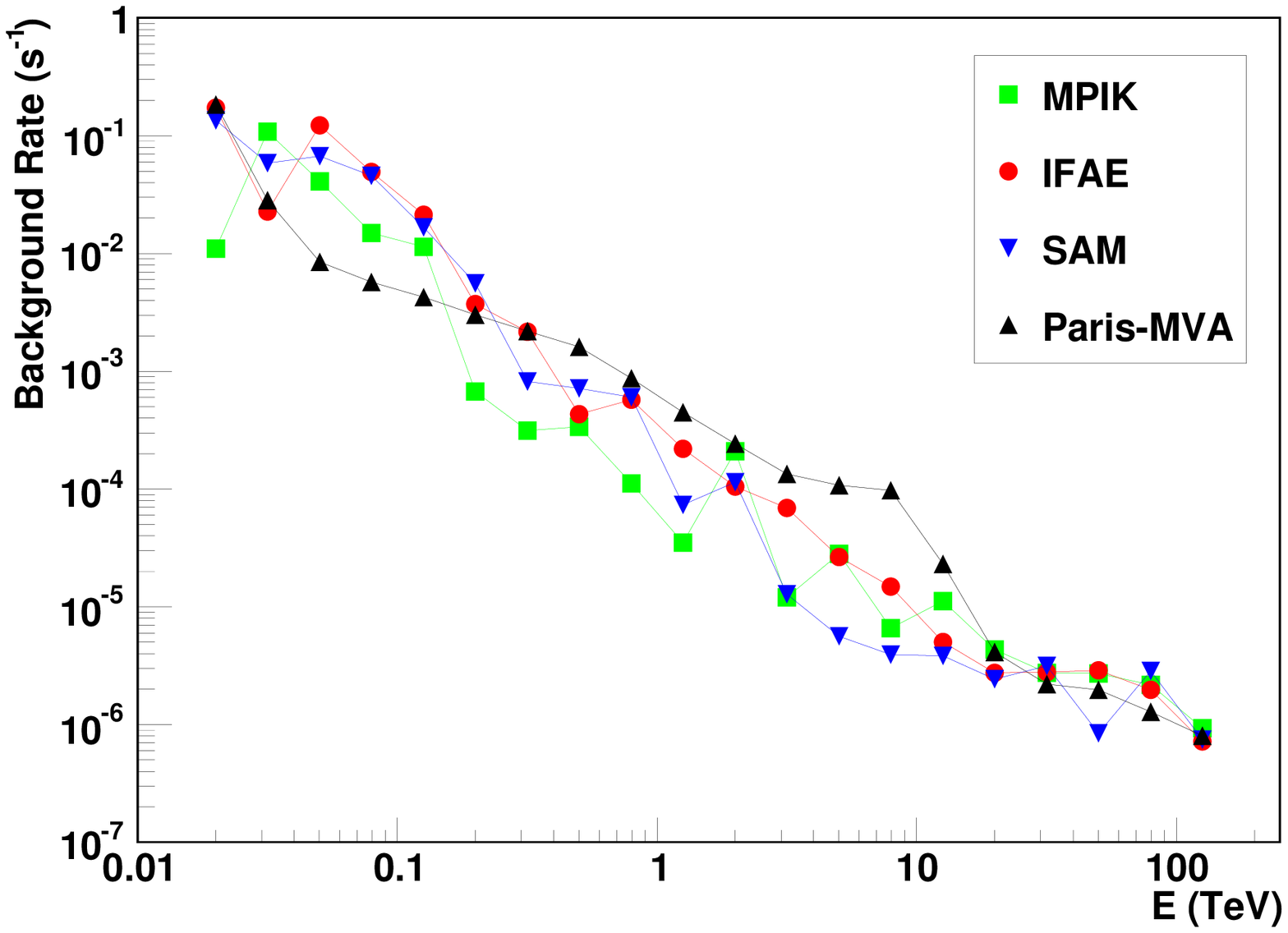}
    \caption[CR background rate]{
      Residual cosmic-ray background rate of candidate array I,
      given as a function of the estimated energy
      for the baseline/MPIK (green squares), IFAE (red circles),
      SAM (blue triangles) and Paris-MVA (black triangles) analyses.
      The differential sensitivities are optimized for an observation time of 50 hours.
    }
    \label{fig7:backgnd}
\end{figure}

In the range of 30~GeV to 3~TeV, the Paris-MVA analysis improves
on the sensitivity of the baseline analysis, by up to a factor of $3.5$.
We note that the Paris-MVA analysis generally maintains a 
large effective area with respect to the baseline analysis,
although it also yields the largest cosmic-ray background rate for energies 
above 300~GeV.
The SAM analysis provides the best performance at multi-TeV energies, 
improving on baseline result by factors as large as $2.0$, with effective
areas not quite as large as Paris-MVA but compensated by low
background rates.


\subsection{The angular and energy resolutions}

The $\gamma$-ray events used to determine the angular and energy resolutions 
are those that pass the cosmic-ray background rejection cuts, which are used 
to obtain the respective differential sensitivity curves in 
Figure~\ref{fig7:diffsens}.
The $r68$ angular resolution, is defined as the angular radius from a 
point-like source that contains $68\%$ of the events.
This is shown as a function of the estimated energy for the case of 
candidate array I in Figure \ref{fig7:angres}.
It varies, depending on the analysis, from $0.2^{\circ} - 0.5^{\circ}$ 
at $\sim20$~GeV to $0.02^{\circ} - 0.03^{\circ}$ at $\sim125$~TeV.
The different shower direction reconstruction methods and various 
cosmic-ray background rejection cuts
lead to the noticeable spread in $r68$ values below $\sim 1$~TeV between 
the alternative analyses.

\begin{figure}
    \includefigure{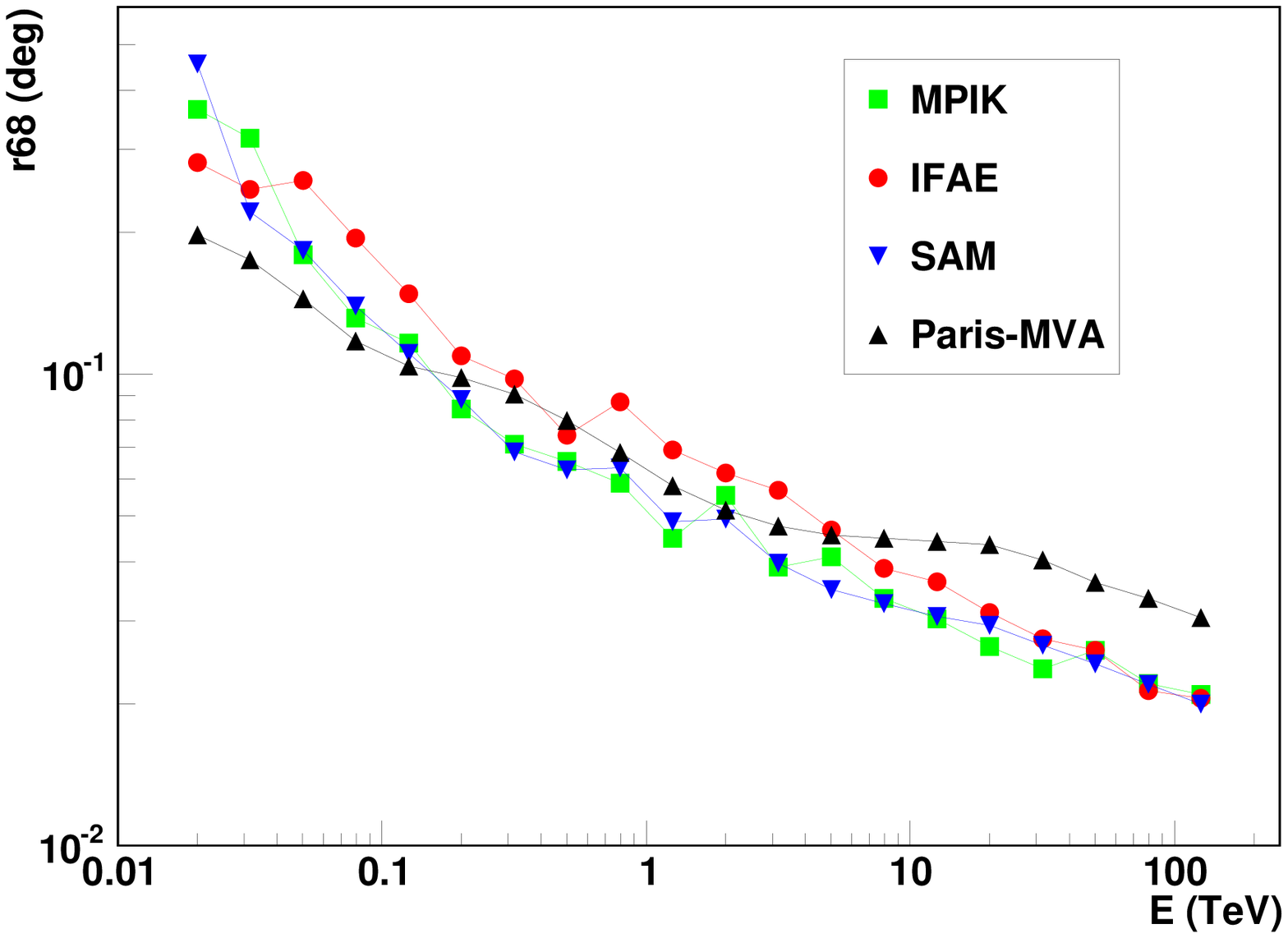}
    \caption[Angular resolution in alternate analyses]{
     Angular resolution (radius of $68\%$ event containment, $r68$) of candidate array I,
     given as a function of the estimated energy
     for the baseline/MPIK (green squares), IFAE (red circles),
     SAM (blue triangles) and Paris-MVA (black triangles) analyses.
     The differential sensitivities are optimized for an observation time of 50 hours.
   }
    \label{fig7:angres}
\end{figure}

A similar argument may be used to explain the differences
between the alternative energy resolutions of the candidate array I,
although here the methods used to determine the estimated energy also differ, as detailed in Section~\ref{sec:alt_methods}.
We define the term `energy resolution' to be the r.m.s.\ of the distribution 
$E_{\mathrm{est}}/E_{\mathrm{true}}$,
where $E_{\mathrm{est}}$ is the estimated energy and $E_{\mathrm{true}}$ is the true energy.
Figure \ref{fig7:eres} shows the energy resolutions as functions of the estimated energy.
The energy resolutions range between $0.3-0.50$ at around 20 GeV to $\sim 0.05$ at around 8 TeV.
The poor energy resolution of Paris-MVA at high energies is being worked on.

\begin{figure}
    \includefigure{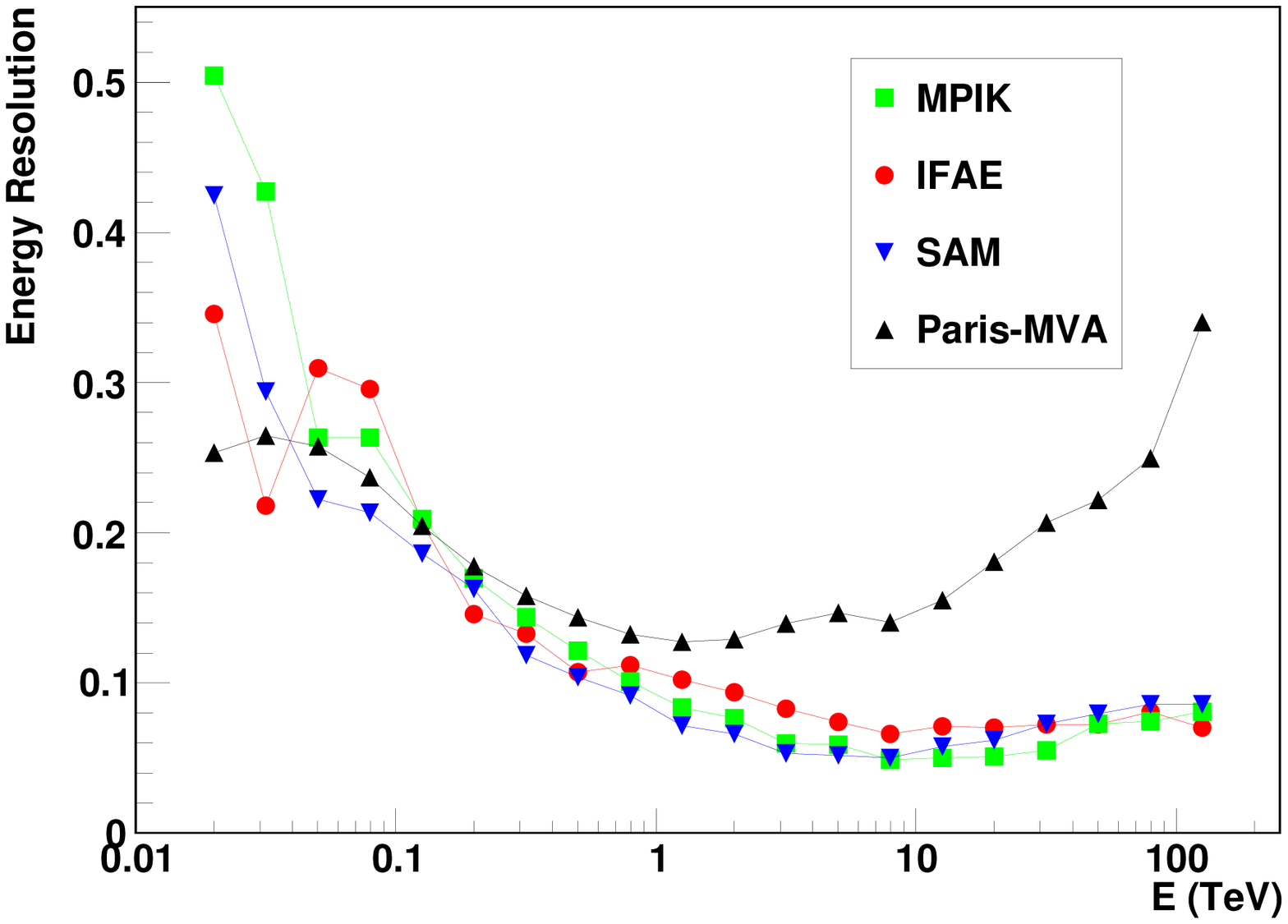}
    \caption[Energy resolution in alternate analyses]{
     Energy resolution of candidate array I, defined as the r.m.s.\ 
     of the distribution of $E_{\mathrm{est}}/E_{\mathrm{true}}$,
     using the baseline/MPIK (green squares), IFAE (red circles),
     SAM (blue triangles) and Paris-MVA (black triangles) analyses.
     $E_{\mathrm{est}}$ is the estimated energy, while $E_{\mathrm{true}}$ is the true energy.
     The energy resolution is given as a function of the estimated energy.
     The differential sensitivities are optimized for an observation time of 50 hours.
   }
    \label{fig7:eres}
\end{figure}


\subsection{Observations under partial moon light}
\label{sec:moon-light}

The standard observing conditions of ground-based Cherenkov telescopes correspond
to clear sky, with both Sun and Moon below the horizon, and lead typically to an average
of 1650~h/yr of possible dark observation time.
Allowing observations under moderate moonlight
conditions increases the total observation time by up to 30\%. This can be of
crucial importance in case of transient phenomena, such as flares of AGNs,
phase-related activities as for binary systems, or GRBs. The MAGIC and VERITAS
experiments both routinely perform observations under moonlight conditions. MC
simulations with a NSB 4.5 times higher to the one commonly in use have been
performed. This corresponds to nights with the Moon above the horizon,
approximately illuminated at 60\%, although often moon-time observations can
be carried out at  less enhanced NSB. Trigger thresholds have been adjusted to
obtain manageable trigger rates -- while in reality also the PMT gain may get
adjusted to reduce additional PMT aging. The simulations have been processed through the
entire analysis chain, as discussed in the preceding sections, with somehow
different cuts, particularly in the image cleaning steps to account for the
higher NSB noise levels.

The final results are similar to those obtained for dark sky conditions
\cite{ICRC2011}, and hardly affect the performances above 1~TeV. Indeed, as
expected, the moonlight observing conditions mainly affect the low part of CTA
energy range. In this part, due to higher noise levels, and depending on the
array layout, the energy thresholds are generally a factor of two higher and the
sensitivities might be ten times worse. However, in the core of CTA energy range
around 1~TeV, not only the sensitivity but also the angular and energy
resolutions are quite compatible with the results of dark sky observations.


\subsection{High-altitude sites}
\label{sec:high-altitude}

The altitude of 2000~m has been assumed for most of the CTA simulations
but the possible performance advantages (and disadvantages) of high-altitude
sites have been investigated as well. First comparisons were carried out
with arrays of 9 large telescopes at 2000, 3500, and 5000~m altitude
\cite{ber2008a,cta2010}. High-altitude sites result in lower energy
thresholds, mainly because the Cherenkov light gets less diluted when
reaching ground. For the same reason, smaller telescope separations
are needed, resulting in lower effective areas at medium to high
energies. Images are also seen at larger distances from the shower
direction, which may require a larger camera f.o.v.
At the 5000~m altitude, gamma-hadron separation was found to
suffer from too many Cherenkov-emitting particles in $\gamma$-ray showers
reaching ground level, resulting in more irregular $\gamma$-ray images.

For the latter reason, as well as due to technological and cost implications of
a high-altitude site, more detailed simulations of high-altitude
sites were limited to 3700~m, for the same telescope configuration
as in production-1. Neither were telescope separations optimized
for the higher altitude nor were cost implications included in the
array selection, e.g.\ using array I as for 2000~m.

Initial results of different studies show that the
energy threshold achieved at 3700~m is in general lower by 0.2~dex or more, and
in the low energy domain (E$<$100~GeV), the differential sensitivity achieved is
generally at least a factor two better. At higher energies, the achieved
differential sensitivities of the different layouts envisaged for CTA at high
altitude are compatible with those obtained at 2000~m, perhaps marginally worse. 
However, the energy resolution generally degrades above a few TeV.
Before firm conclusions can be drawn on an optimum site altitude,
separate optimisations of telescope spacings at the different altitudes
are required and cost implications (in telescope design, construction, and 
operation) should be represented in the array selection, for a
performance comparison at fixed cost.

\section { Future directions and hardware }
\label{sec:FutureHardware}

\subsection { Hardware improvements }

The Monte Carlo simulation of the detector was done assuming a very preliminary
design of the CTA telescopes. Both due to the hardware development and the
feedback from the simulations, some of the designs being under discussion
currently differ from the simulated one. Monte Carlo simulations are going on to
understand the improvements and limitations of those modifications.

\subsubsection { Changes in optical design }

The initial simulations assumed a parabolic dish shape for
the LSTs and a Davies-Cotton shape for MSTs and SSTs.
While the parabolic shape has the advantage of negligible
time-spread introduced by the optics, the classical 
Davies-Cotton design has the advantage of a better
PSF at large off-axis angles. While the optics time-spread
of SSTs is always small compared to the shower-intrinsic
spread as well as the PMT transit-time jitter, the time-spread
of Davies-Cotton MSTs is already dominated by the optics.

The MST optics has thus been redesigned and changed to
an intermediate shape, with spherical dish of radius of
curvature $R_c=1.2 f$. Note that Davies-Cotton has $R_c=f$ while
a parabolic dish has a central radius of curvature of $R_c=2f$.
The modified MST optics is still close to a Davies-Cotton
but with a significantly reduced time spread (0.7~ns r.m.s.
instead of 1.0~ns).
The slightly worse off-axis PSF of the intermediate shape is
compensated by a small increase in its focal length
(from 15.6 to 16.0~m).

A similar intermediate shape, but closer to parabolic,
has also been recommended for the LSTs. Only muon rings,
easily recognizable in images, have an intrinsic time spread
short enough that they could take any advantage of the
low time spread of a parabolic dish. Otherwise, both
for gamma-ray showers and background, the intrinsic time spread
of photons imaged into the same pixel has an r.m.s.\ value of 
the order of 0.5 to 1.0~ns in the energy and core distance
range relevant for LSTs. The PMT transit time jitter (or spread) is of
a similar magnitude. Considering that, an intermediate dish
shape with an optical r.m.s.\ time spread of 0.6 ns has been
recommended, which is closer to parabolic than Davies-Cotton
but already offers improvements in off-axis PSF.

\subsubsection { Camera Trigger }

The results presented in previous sections all used a majority trigger logic with 
a low combinatorial factor, requiring that a pixel
plus a number of its direct neighbours must have fired within a
given gate width. None of the current
hardware developments in CTA is focusing on such an algorithm yet, 
although some of the developed options are flexible enough to program it.

There are several trigger designs considered that use the analogue signal from the
PMTs and others that use digitized signals, either fully digitized samples
(from FADCs) or comprising only one or two bits (from fast comparators).

\begin{itemize}
\item{\bf Majority Trigger:} The analogue signal coming from each pixel
is compared to an adjustable 
threshold (by a discriminator or comparator) and then the sum of
discriminator/comparator outputs in a region (either analogue or digital sum, 
basically the number of pixels above 
threshold at the same time) has to exceed an
adjustable multiplicity value to result in camera triggers. 
\item{\bf Analogue Sum Trigger:} The analogue signals from all pixels in a
region are added and then compared to a minimum value to produce the camera
trigger. Before adding the individual pixel signal, they can be clipped, limiting
their maximum value and thus reducing the impact of afterpulses.
\item{\bf Binary Trigger:} The analogue signal from the photosensors is passed
through a comparator at regular time intervals, essentially transforming the camera
image to a binary pattern.
With modern fast and flexible front-end electronics,
complex trigger classification algorithms can be run on-line, processing the 
binary pattern in space and time. With additional thresholds, this trigger
concept can emulate image cleaning algorithms, similar to those used for off-line image
analysis.
\item{\bf Digital Trigger:} The signal coming from the photosensors is
continuously digitized by a FADC and the digital signal is used to take the trigger
decision - which could be either a digital sum trigger with optional clipping 
or a digital majority trigger 
(or a mix of both), depending on the algorithms programmed into FPGAs.
This scheme avoids front-end electronics like discriminators and comparators,
but is currently only 
cost-effective for FADC sampling rates up to about 250 MHz.
\end{itemize}

Trigger simulation tools exist for all of these different schemes.
The key quantities for assessment of any improvement of the trigger performance
should be sensitivity, angular resolution, and energy resolution. 
The triggered images still need to be subjected to a reconstruction, and it is
useless to trigger more images if that cannot help to better reconstruct the shower and
to improve the gamma-hadron separation.
Data analysis techniques are still improving
and there might eventually be methods to recover some fraction of the events currently
discarded -- but at least with current analysis techniques the events barely
triggered tend to have too low amplitude signals to be usable for reconstruction.
The study of the impact of improved trigger schemes in terms of
source sensitivity etc.\ is work in progress.

\subsection { Optimal pixel integration time window }

Most of the detected Cherenkov photons are emitted near the shower maximum
and arrive at an IACT pixel within only 
a few nanoseconds time spread. As a consequence of the short time scale of 
Cherenkov light flashes, there is an optimum integration time which minimizes 
the error in collected pixel signal charge over NSB fluctuations. 
For short integration times, the Cherenkov signal should dominate over NSB.
As the integration time
increases, the shower fluctuations are smoothed out and the relative error on
the integrated charge decreases. Once the bulk of the signal is integrated, a
further increase of the integration time will degrade the accuracy due to the
NSB fluctuations (see Fig.~\ref{fig:IntegrationTime} 
\textit{\ifwithtwocolumns (top)\else (left)\fi}).

\begin{figure}[htbp]
\ifwithtwocolumns
\includefigure{plot4-sec8}
\includefigure{Fig20_RightPlot_PoissonSep_c}
\else
\centering
\includegraphics[width=0.49\columnwidth]{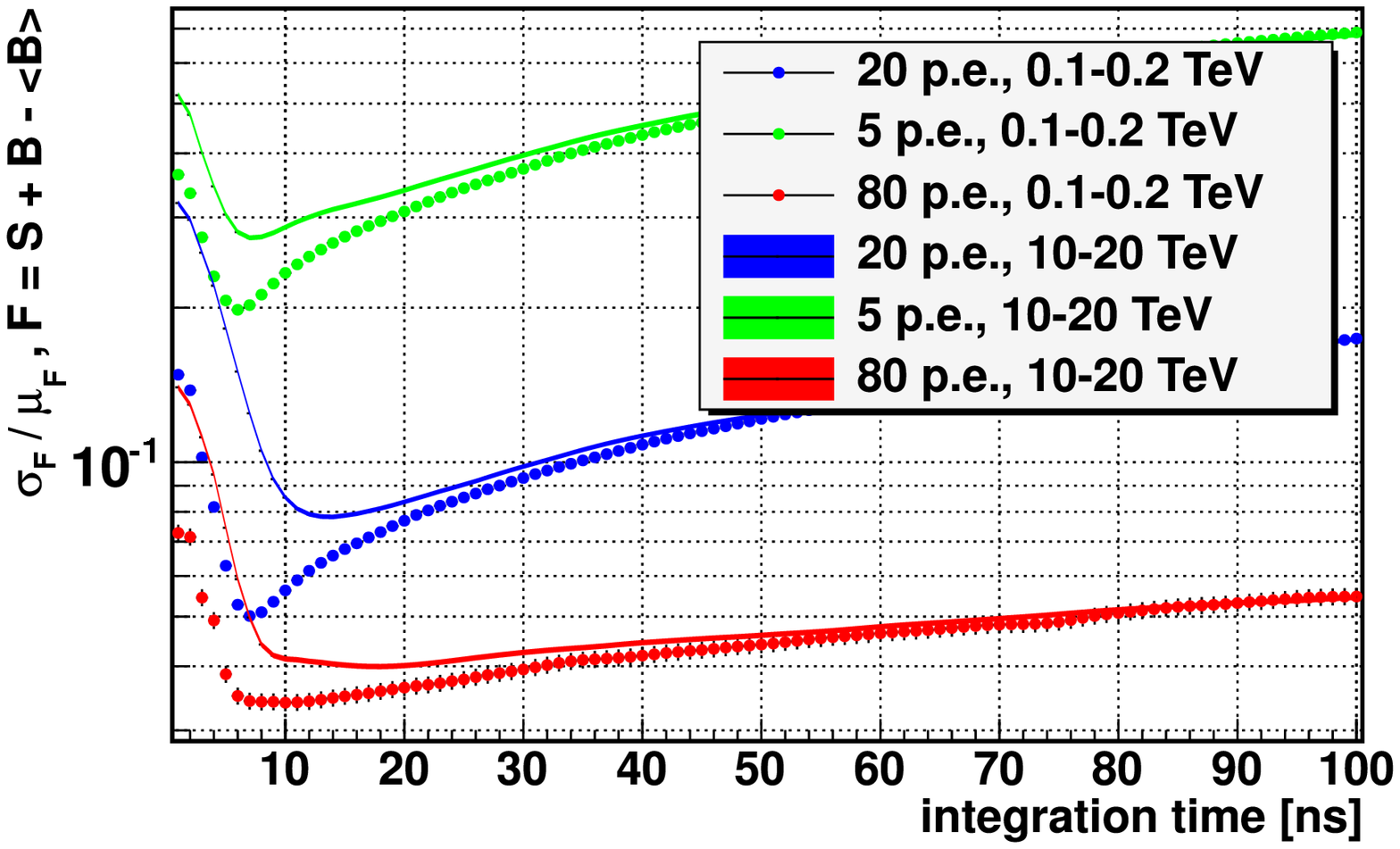}
\includegraphics[width=0.49\columnwidth]{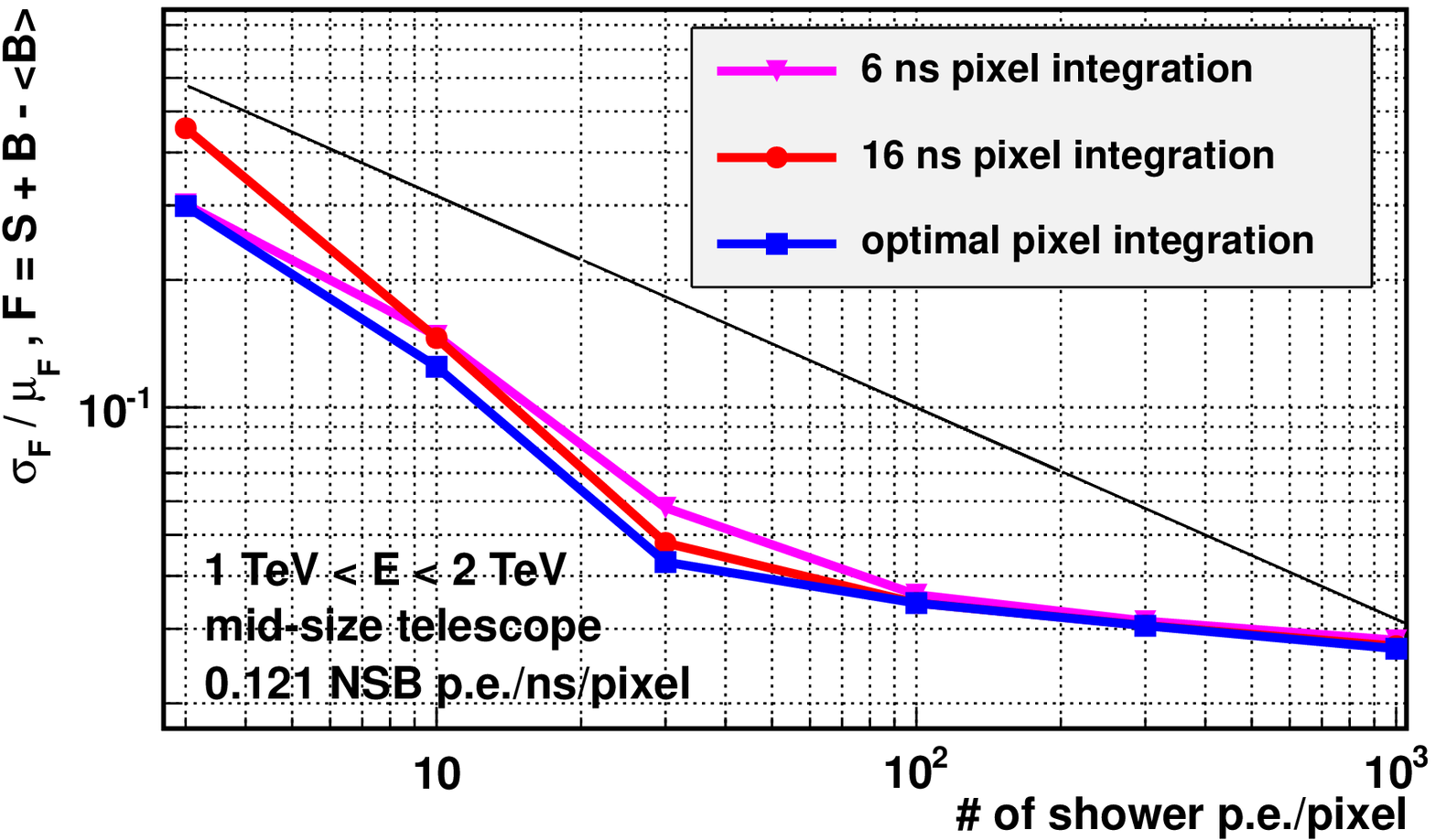}
\fi
\caption[Charge error versus integration time]{
Relative error on the integrated charge by mid-size telescope pixels:
\textit{\ifwithtwocolumns (top)\else (left)\fi} 
as a function of the integration time, for the indicated signal p.e.
charges, and \textit{\ifwithtwocolumns (bottom)\else (right)\fi} 
as a function of the number of shower p.e. per pixel,
for two fixed and one dynamic integration windows (see text). The error represents
the ratio standard deviation/mean of the distribution of the functional $\mathrm{F = S + B -
{\langle B \rangle}}$, where S, B and $\mathrm{\langle B \rangle}$ are signal, NSB,
and the average  NSB (0.121 p.e./ns/pixel, dark sky) contributions.
Note that the unavoidable $1/\sqrt{N_{\rm pe}}$ Poisson fluctuations in the number 
of registered p.e.s are not included
but shown separately (straight black line) and 
dominate in all cases shown here.
}
\label{fig:IntegrationTime}
\end{figure}

Even at high energies, most of the Cherenkov light  can be collected within a
$\sim$15-20 ns \textit{per pixel} time window, see Figures \ref{fig:IntegrationTime} 
and~\ref{fig:sigovernsb}. The
dynamic range of the pixel charges in the event shown in Fig.~\ref{fig:sigovernsb}
is up to $\sim 10^4$~p.e. As in most gamma-ray shower images, the largest
signal amplitudes correspond to the shower maximum. The
correlation between the optimal integration time and the signal amplitude in a pixel
is clearly visible. A dynamic integration window, i.e.\ the one where the duration is
varied as a function of signal amplitude to provide the best pixel charge resolution,
could be an improvement with respect to a fixed duration integration time (see
Fig.~\ref{fig:IntegrationTime} \textit{\ifwithtwocolumns (bottom)\else (right)\fi}).
Under conditions of dark sky the overall improvements are very small, since
Poisson fluctuations dominate, but under partial moon light conditions
more significant improvements can be envisaged.
Whether the additional effort (and cost) of its technical
implementation and calibration is worth this effort needs to be studied.

\begin{figure}[htbp]
\ifwithtwocolumns
\includefigure{Fig21_LeftPlot}
\includefigure{Fig21_RightPlot}
\else
\centerline{
{\includegraphics[width=0.468\columnwidth]{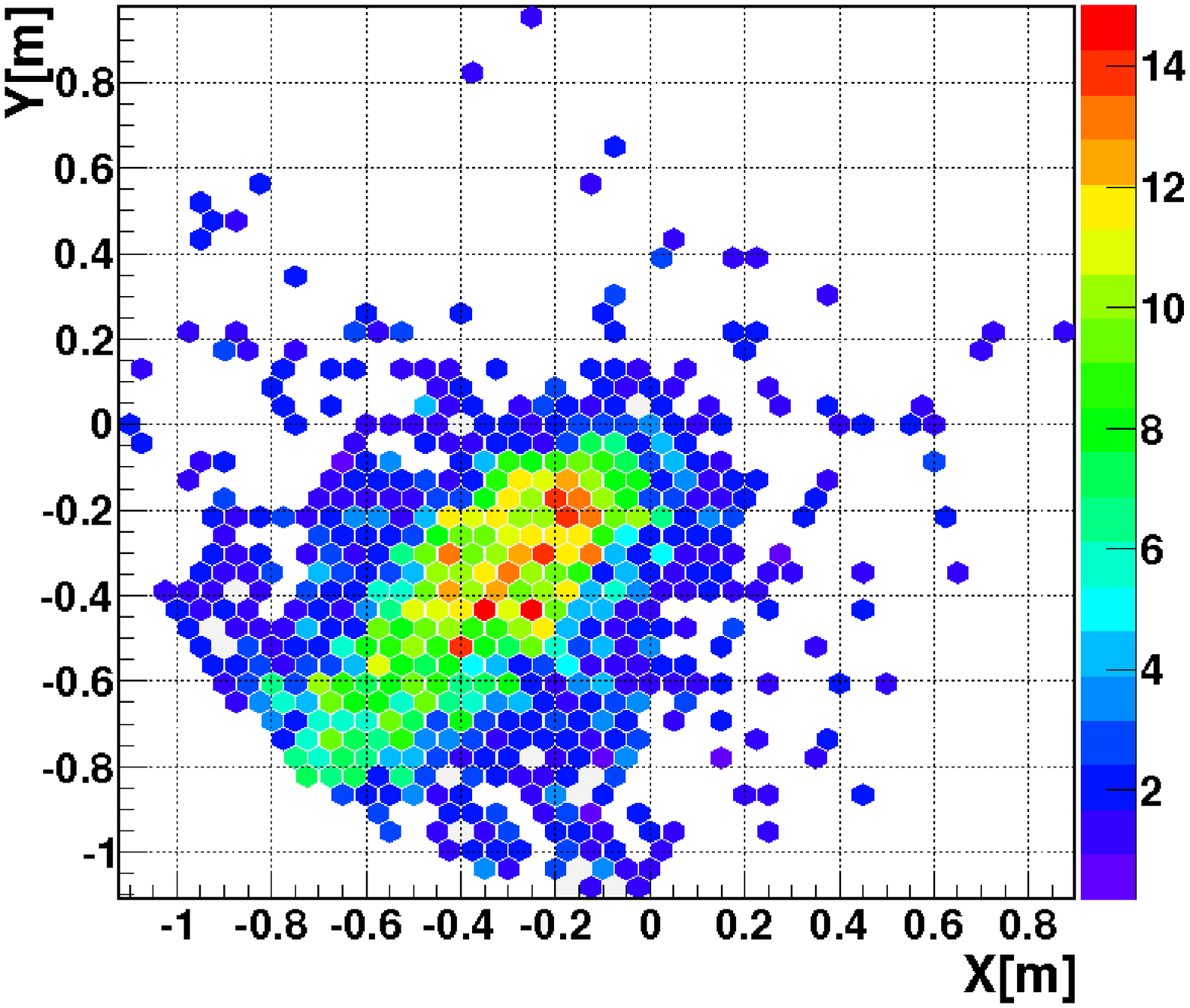}
\label{fig:mbwidth}}
{\includegraphics[width=0.461\columnwidth]{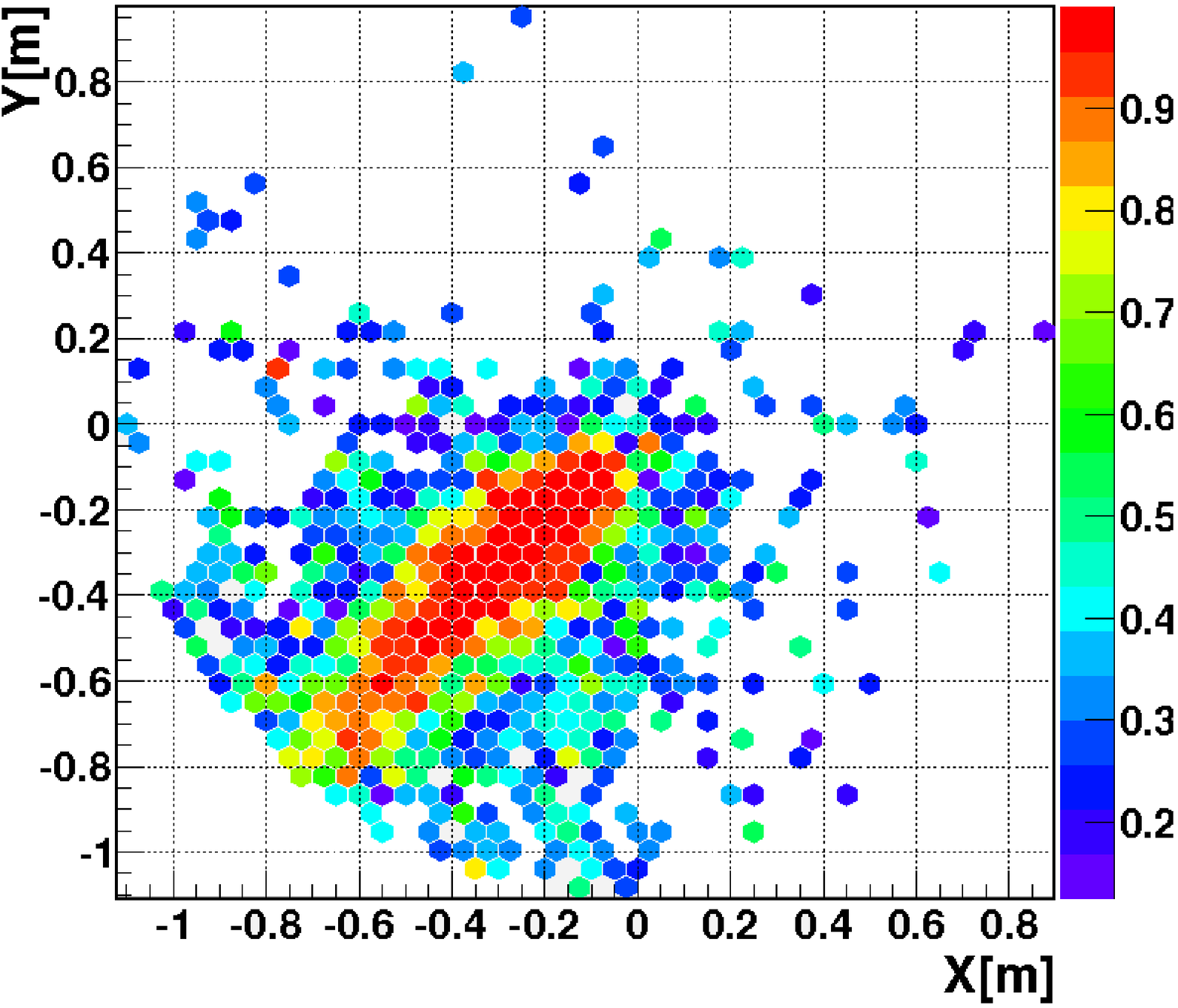}
\label{mbdq}}
}
\fi
\caption[Optimal integration time on image]{
\textit{\ifwithtwocolumns (top)\else (left)\fi} Focal plane distribution of the optimal pixel integration time values (ns), for a 49 TeV $\gamma$-ray event with 133 m impact parameter.
\textit{\ifwithtwocolumns (bottom)\else (right)\fi} For the same event, distribution of the fractions integrated/total signal charge. Only pixels with an actual Cherenkov signal in the simulation are shown.}
\label{fig:sigovernsb}
\end{figure}

\subsection{ Readout strategy }

As it has been mentioned, Cherenkov photons do not all reach a telescope at
the same time. For instance, the time spread  for photons coming from a 10
TeV shower can be, in extreme cases, as large as 200~ns 
while in a  single pixel is at least one order of magnitude smaller. 
Actually, pixels receiving Cherenkov light only from a single particle
may see all photons arriving within a fraction of a nanosecond while
pixels with photons from multiple particles have most of these photons
arriving within about one to a few nanoseconds - while the whole
camera typically sees photons spread out from a few nanoseconds to
tens of nanoseconds, for large off-axis showers at large core distances
seen in wide-field telescopes even beyond 100 ns.

The current  generation of Cherenkov detectors are using analogue memories and 
reading the signal at the same time for all pixels. Hence,  either one has
to read out a large buffer section of the analogue memories or a significant
amount of the signal is not  recorded for some showers. The former would
increase the dead-time and the latter would reduce the collection area. A
compromise strategy is to have a so-called "sliding" readout window. A short
readout time is used for all pixels, but the start  of the readout is not
made at the same time for all pixels. It follows the signal propagation
within a camera. Possible real implementations of this readout scheme are
being developed in the CTA consortium and these require the transmission
of  information of triggered pixels to the readout device of these associated
pixels. These strategy will be implemented in the  future full simulations to
quantify their impact on the global performance of CTA.

\subsection{Simulation of dual-mirror telescopes}

While the simulation tools initially were restricted to
telescopes with a single (segmented) reflector, they
have been extended to include telescopes with
secondary optics of the Schwarzschild-Couder (SC) design 
\cite{SCoptics}, of which two types are under
development for CTA.

\subsubsection { Small Size Telescopes with secondary optics }

Even though only Small Size Telescopes with Davies-Cotton (DC) optics have 
been simulated on a large scale so far (Table 1), 
proposals have also been made for the use of even smaller telescopes
of the 3.5 to 4~m class,
using secondary optics following the SC design 
to cover the high energy regime of the array. 
Such a telescope could be
significantly less expensive due to the lower price per pixel for the 
photosensors available at
the reduced plate scale of the SC optics, allowing more SSTs to be built,
resulting in both improved angular resolution and effective area at
high energies.

Of course with such an untested telescope design detailed simulations
must be performed to try and quantify their effect on the performance
of the array. Preliminary simulations have shown improvements can be
made to the array effective area, while having little effect on the
angular resolution. Larger scale simulations are planned,
aimed at accurately determining the telescope performance
(including background discrimination ability), as well as finding the
optimum array layout. 

In addition to these large scale simulations, simulations of specific
aspects of the SC-SST designs are required, for example to quantify the
performance of different photosensor/electronics combinations, such as 
multi-anode PMTs (MAPMTs) and silicon photomultipliers (SiPMs).

\subsubsection { Schwarzschild-Couder Mid Size Telescopes }
\label{sec:sc-mst}

Continuing work begun by the AGIS Collaboration, which joined with CTA in
September 2010, the US groups are leading an effort to develop two-mirror
mid-size telescopes of SC optical design, and
planning to extend the initial DC-MST array of about 20 telescopes by an additional
36 SC-MSTs. With O(60) telescopes, the MST array will fully enter the
event-containment regime, where the effective area within the array is
comparable to, or greater than, the effective area around the array's edges.
This is an important advance over current arrays of 2-4 telescopes, for which
the effective area is dominated by events that land outside the array (for
example, only 5\% of 1 TeV gamma rays land within the $10^4$~m$^2$ footprint of
the VERITAS array). Contained events are much better sampled, providing
improved background rejection (more likely to pick up anisotropies in hadronic
showers), improved angular resolution (triangulation of the shower direction
is much more effective when the shower is viewed from several widely spaced
azimuthal angles), and reduced energy threshold (containment implies that the
shower's impact distance to the several nearest telescopes will be less than
the 100-200 m telescope spacing)~\cite{maier2009,cta2010}.

The baseline SC-MST optical design has a  $\sim$9.5~m diameter primary mirror
(4.4 m central hole), 5.4-m diameter secondary mirror, and a 5.6 m focal
length, providing $\sim$50 m$^2$ of effective light collecting area. The SC design
offers several advantages over the DC design which become especially important
for a large array. The SC optical design corrects for spherical and comatic
aberrations and is optimized to minimize astigmatism, keeping the optical PSF
smaller than the size of a pixel out to 4--$5^\circ$\ off axis and providing
a short focal length compared to DC-MSTs with a similar field of view.
Increasing the field of view increases the telescope multiplicity of each
event \cite{bugaev2009}. The demagnifying secondary mirror reduces the plate
scale of the focal plane:  an $8^\circ$\ field of view requires only a
focal plane of 0.8~m diameter, providing a large reduction in per-channel costs
for focal-plane instrumentation by enabling the use of, for example,
64-channel multi-anode PMTs. The small plate scale also makes it economical to
reduce the angular size of each pixel from the $\sim0.15^\circ$\ used in
current-generation telescopes to $\sim0.07^\circ$, which is expected to
significantly improve angular resolution \cite{bugaev2009}. The SC design also
has no wavefront distortions, allowing tighter requirements on trigger timing,
which ought to reduce the rate of accidental triggers at a given threshold and
may allow operation at a lower energy threshold. Finally, the small plate
scale and opportunity to use multi-channel photodetectors improves the
modularity and serviceability of the focal plane instrumentation and data
acquisition electronics.

Several challenges are introduced by the SC design, and a major thrust of the
simulations effort will be to study trade-offs between cost and performance to
refine the specifications for the SC telescopes. Aspects under study include
\begin{itemize}
\renewcommand{\labelitemi}{-}
\item the field of view and spacing of the telescopes, which impacts the number of
channels of focal plane instrumentation required and the quality of the
optical PSF at the edge of the field,
\item the quality of the optical PSF across the field of view and angular size of
the pixels, which again speaks to the number of channels required, as well as
requirements on the mirror alignment precision and the optical quality of the
aspheric mirror segments,
\item the mechanics of the focal plane, including the accuracy to which pixels
need to be placed across the curved focal plane, tolerances for optical cross
talk and dead spaces between pixels, and whether light concentrators provide a
benefit,
\item the impact of vignetting on the optical design specifications and
performance of the analysis of Cherenkov images near the edge of the field of
view,
\item requirements on the slew speed of the telescopes, driven by gamma-ray
burst follow-up,
\item studies of various options for triggering electronics to evaluate
trade-offs between maximizing rejection of accidental and cosmic-ray triggers,
maximizing the low-energy effective area, and minimizing costs. These studies
will also impact the specifications for the data acquisition electronics, in
particular setting the dead-time and throughput specifications,
\item an evaluation of the overall impact of adding 36 SC-MSTs to the 23 planned
DC-MSTs on the sensitivity, angular resolution, and energy threshold of the
MST array.
\end{itemize}

\section { Conclusions and outlook }
\label{sec:conclusion}

We could demonstrate that current shower and
telescope simulation methods agree well with
each other and with measured data, verifying
the simulation tools in use.
After first large-scale simulations have shown that the initial
goals for the CTA sensitivity are quite
reasonable over most of the anticipated energy range,
the current simulation data sets have resulted
in a number of array layouts which can satisfy the
expectations for most physics goals and in a number
of layouts more focused on individual science cases.
The benefits of the huge improvement in
performance by CTA with respect to the current generation
of instruments are demonstrated in other papers
of this journal issue.
Only at the very lowest energies, the initial
assumptions on a CTA sensitivity curve could not
be fully met, due to fluctuations in hadron showers
resulting in gamma-like background events and because the
background distribution over the f.o.v. cannot be
known perfectly (inclusion of background systematics
in the analysis). In its core energy range, the
initial expectations for CTA can be met if not
surpassed. At the highest energies, the development
of more cost-effective small telescopes may well
result in effective areas well exceeding initial plans.

The analysis methods for an instrument like CTA are
still under development, having achieved quite some
improvements over the baseline method which is based
on only the traditional Hillas parameters. While this
development will continue it is important to see that
a CTA layout optimal with one analysis method is also
close to optimal with other methods. Our iterative
procedure in optimizing the CTA layout and configuration
can thus continue with a superset of near-optimal
layouts, plus perhaps some borderline cases, in its
next round. Together with improved cost estimates,
this should provide the basis for an optimal CTA
performance for almost any of the CTA astrophysics
or fundamental physics goals.

\section*{Acknowledgements}

We gratefully acknowledge support from the following agencies and organisations:
Ministerio de Ciencia, Tecnolog\'ia e Innovaci\'on Productiva (MinCyT),
Comisi\'on Nacional de Energ\'ia At\'omica (CNEA) and Consejo Nacional  de
Investigaciones Cient\'ificas y T\'ecnicas (CONICET) Argentina; State Committee
of Science of Armenia; Ministry for Research, CNRS-INSU and CNRS-IN2P3,
Irfu-CEA, ANR, France; Max Planck Society, BMBF, DESY, Helmholtz Association,
Germany; MIUR, Italy; Netherlands Research School for Astronomy (NOVA),
Netherlands Organization for Scientific Research (NWO); Ministry of Science and
Higher Education and the National Centre for Research and Development, Poland;
MICINN support through the National R+D+I, CDTI funding plans and the CPAN and
MultiDark Consolider-Ingenio 2010 programme, Spain; Swedish Research Council,
Royal Swedish Academy of Sciences financed, Sweden; Swiss National Science
Foundation (SNSF), Switzerland; Leverhulme Trust, Royal Society, Science and
Technologies Facilities Council, Durham University, UK; National Science
Foundation, Department of Energy, Argonne National Laboratory, University of
California, University of Chicago, Iowa State University, Institute for Nuclear
and Particle Astrophysics (INPAC-MRPI program), Washington University McDonnell 
Center for the Space Sciences, USA. 
The research leading to these results has received funding from the European 
Union's Seventh Framework Programme ([FP7/2007-2013] [FP7/2007-2011]) under 
grant agreement no.~262053.



\bibliographystyle{model1-num-names}








\else





\fi

\end{document}